%% file: arxiv-final.tex
\documentclass[conference]{IEEEtran}
\IEEEoverridecommandlockouts
\usepackage{cite}
\usepackage{amsmath,amssymb,amsfonts}
\usepackage{algorithmic}
\usepackage{graphicx}
\usepackage{textcomp}
\usepackage{xcolor}
\def\BibTeX{{\rm B\kern-.05em{\sc i\kern-.025em b}\kern-.08em
    T\kern-.1667em\lower.7ex\hbox{E}\kern-.125emX}}

\usepackage{subcaption}
\usepackage{microtype}
\usepackage{hyperref}
\usepackage{url}
\usepackage{booktabs}

\usepackage{lineno}

\definecolor{darkblue}{rgb}{0, 0, 0.5}
\hypersetup{colorlinks=true, citecolor=darkblue, linkcolor=darkblue, urlcolor=darkblue}

\newcommand{\Ds}{\mathcal{D}_{\text{sanitized}}}
\newcommand{\Do}{\mathcal{D}_{\text{original}}}

\usepackage{microtype}
\usepackage{graphicx}
\usepackage{booktabs} %

\usepackage{rotating}
\usepackage{adjustbox}

\usepackage{amsmath}
\usepackage{amssymb}
\usepackage{mathtools}
\usepackage{amsthm}

\usepackage{makecell}
\usepackage{multirow}
\usepackage{tcolorbox}
\usepackage{algorithm2e}
\usepackage{listings}

\usepackage{hyperref}

\usepackage{longtable}

\usepackage{booktabs} 
\usepackage{setspace}

\usepackage[numbers,sort&compress]{natbib}

\usepackage{cuted}

\usepackage{multicol}
\usepackage{listings}
\usepackage[hybrid]{markdown}

\begin{document}

\title{A False Sense of Privacy:  Evaluating Textual Data Sanitization Beyond Surface-level Privacy Leakage}

\author{
\IEEEauthorblockN{Rui Xin\textsuperscript{$*$}\thanks{\textsuperscript{$*$}Equal Contribution}}
\IEEEauthorblockA{\textit{University of Washington}\\
rx31@cs.washington.edu}
\and
\IEEEauthorblockN{Niloofar Mireshghallah\textsuperscript{$*$}}
\IEEEauthorblockA{\textit{Carnegie Mellon University}\\
niloofar@cmu.edu}
\and
\IEEEauthorblockN{Shuyue Stella Li}
\IEEEauthorblockA{\textit{University of Washington}}
\and
\IEEEauthorblockN{Michael Duan}
\IEEEauthorblockA{\textit{University of Southern California}}
\and
\IEEEauthorblockN{Hyunwoo Kim}
\IEEEauthorblockA{\textit{NVIDIA}}
\and
\IEEEauthorblockN{Yejin Choi}
\IEEEauthorblockA{\textit{Stanford University}}
\and
\IEEEauthorblockN{Yulia Tsvetkov}
\IEEEauthorblockA{\textit{University of Washington}}
\and
\IEEEauthorblockN{Sewoong Oh}
\IEEEauthorblockA{\textit{University of Washington}}
\and
\IEEEauthorblockN{Pang Wei Koh}
\IEEEauthorblockA{\textit{University of Washington}}
}

\maketitle

\begin{abstract}

Sanitizing sensitive text data typically involves removing personally identifiable information (PII) or generating synthetic data  
under the assumption that these methods adequately protect privacy; however, their effectiveness 
is often only assessed by
measuring the leakage of explicit identifiers but ignoring nuanced textual markers that can lead to re-identification.
We challenge the above illusion of privacy by proposing a new framework that evaluates re-identification attacks to quantify individual privacy risks upon data release. 
Our approach shows that seemingly innocuous auxiliary information---such as 
routine social activities---can be used to infer sensitive attributes like age or substance use history from sanitized data. For instance, we demonstrate that Azure's commercial PII removal tool fails to protect 74\% of information in the MedQA dataset. Although differential privacy mitigates these risks to some extent, it significantly reduces the utility of the sanitized text for downstream tasks.
Our findings indicate that current sanitization techniques offer a \textit{false sense of privacy}, highlighting the need for more robust methods that protect against semantic-level information leakage.\footnote{Code is available at \href{https://github.com/ruixin31/false-sense-privacy}{https://github.com/ruixin31/false-sense-privacy}}

\end{abstract}


\section{Introduction}

It is critical to protect user and patient privacy when sharing data for research and commercial collaborations \citep{FDS20,mcmahan2017communication}.
When not properly handled, sensitive data---personal identifiers, location traces, behavioral patterns, etc.---can be exposed through re-identification attacks.
During such attacks, adversaries use auxiliary information to analyze released data and infer information about individuals despite sanitization efforts by the data publisher.
Re-identification attacks have proven effective for structured datasets \citep{narayanan2006break}, where identifiers are clearly defined in tabular format.
This effectiveness establishes them as a key method for evaluating data privacy, an approach similar to those found in statistical disclosure control (SDC) guidelines used by the US Census Bureau \citep{abowd20232010}.

Large language models (LLMs) bring new attention to privacy concerns in \emph{unstructured} textual data \citep{yan2024protecting}.
Beyond explicit identifiers, such as personally identifiable information (PII) like names and addresses, text often contains contextual details that can indirectly reveal identity in ways not typically present in structured datasets.
\textit{We hypothesize that re-identification risk remains high with textual data even after PII removal.}
For example, consider Alice's record in the  sanitized medical dataset shown in Figure~\ref{fig:overview},
where all patient information has been de-identified and PIIs removed.
An adversary might possess auxiliary information about Alice, such as her regular social activities or recent job loss,
obtainable through personal knowledge or public sources like social media.
By exploiting similarities between this external information and entries in the sanitized dataset ~\citep{ganta2008composition}, the adversary could re-identify Alice's record
and then infer sensitive attributes, such as her mental health status or substance use.

Developing appropriate metrics to quantify such re-identification vulnerability in unstructured textual data
presents unique challenges. 
Implementing early theoretical frameworks like adapted k-anonymity \citep{venkatesan2008efficient,lison-etal-2021-anonymisation} and mutual-information approaches \citep{6040853} remains difficult due to challenges in establishing consistent quasi-identifiers and measuring mutual information.
Additionally, standard re-identification attacks assume a one-to-one mapping between original and sanitized documents, which is not available for dataset level sanitization methods such as data synthesis.
Moreover, many 
privacy evaluations rely on lexical metrics such as exact matching \citep{pilan2022text,boutet2025anonymization,carlini2019secret,carlini2022quantifying} or ROUGE \citep{huang2023privacy}, 
overlooking the fact that the same information can be expressed in different words.

\definecolor{linking}{HTML}{BEE7F4}
\begin{figure*}[t]
	\centering
	\includegraphics[width=0.99\linewidth]{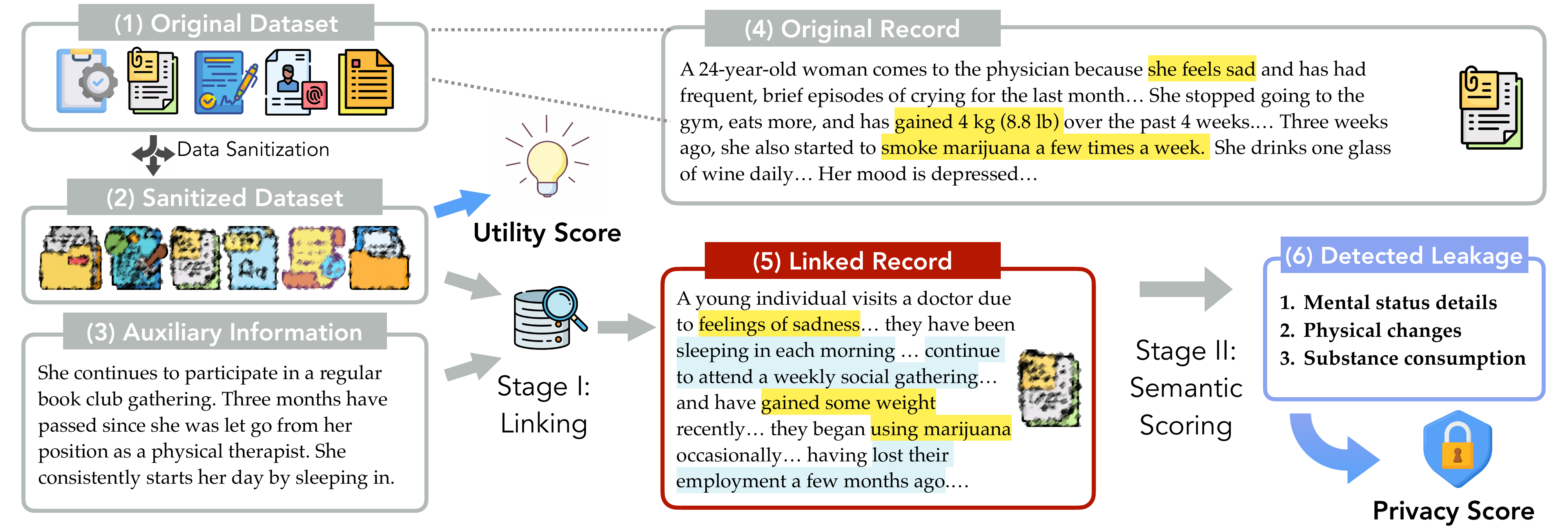}\vspace{-4mm}
 \caption{\setlength{\fboxsep}{0pt} Our privacy evaluation framework overview.
 For a given sanitization method, we obtain the (2) sanitized dataset by applying the method to the (1) original dataset. 
 In the \textbf{linking} stage, we use (3) auxiliary information to find potential matches in the sanitized dataset using a semantic retriever to obtain the (5) linked record.
 Next, in the \textbf{semantic scoring} stage, we analyze the linked record against the corresponding (4) original record to identify \emph{semantic} information leakage.
 In the linked record, text highlighted in \colorbox{yellow}{\strut yellow} indicates detected leakage, and in 
 \colorbox{linking}{\strut cyan} indicates content used for the linking process.
 The framework calculates \textit{utility scores} to measure the practical value of the sanitized dataset and a \textit{privacy score} based on the detected information leakage. 
 }\vspace{-3mm}
\label{fig:overview}
\end{figure*}

To overcome these challenges,
we introduce a new framework for unstructured text that quantifies how much information about an individual can be inferred from sanitized data when combined with auxiliary information \citep{ganta2008composition}. 
Our framework provides a \textit{lower bound} on information leakage, where a low score indicates privacy failure, though a high score does not certify safety against all possible attacks. 
The framework employs a two-stage process (Figure~\ref{fig:overview}).
The first stage, \textbf{linking}, uses a semantic retriever to link auxiliary information with sanitized records.
The second stage, \textbf{semantic scoring}, measures information leakage by comparing the linked sanitized record to the original private data at the atomic claim level.
Simultaneously, we evaluate data utility to assess the tradeoff between privacy and utility.

We evaluate various state-of-the-art sanitization methods on two real-world datasets: MedQA \citep{jin2021disease}, 
with medical notes, and 
WildChat \citep{zhao2024wildchat}, 
with AI-human dialogues 
\citep{mireshghallah2024trust}. 
We compare: %
 (1) \textit{identifier removal techniques}, including commercial PII removal, LLM-based anonymizers \citep{staab2024large}, and sensitive span detection \citep{dou2023reducing}, and
 (2) \textit{data synthesis methods} that use GPT-2 fine-tuned on private data, with and without differential privacy \citep{yue2023synthetic}. 

\textit{We find that current dataset sanitization methods for text data often provide a false sense of privacy}. 
Specifically:  %
(1) State-of-the-art PII removal methods operate at a surface level and still exhibit significant leakage, with 74\% of original information still inferable.
(2) Without differential privacy, synthesized data still exhibits leakage, with 48\% of the information re-identifiable. 
(3) Differentially private (DP) synthesis methods provide the strongest privacy protections but can significantly reduce utility, particularly for complex tasks.
On MedQA, DP-synthesized medical notes have lower utility than the degenerate baseline that removed the original notes. 
Additionally, text quality dropped by 36\%, as evaluated by GPT-4o.

Our evaluation framework establishes that current approaches leave sensitive data vulnerable to re-identification attacks. These results highlight the necessity of developing methods that protect privacy 
beyond surface-level measures and obvious identifiers and of moving toward semantic evaluation of privacy rather than relying solely on lexical assessment.

\section{Related Work}

\noindent\textbf{Privacy evaluations of dataset disclosure.} 
Evaluating privacy prior to dataset release has been a longstanding practice in the statistical disclosure control (SDC) field \citep{hundepool2012statistical}. This practice spans various domains, including legal, technical, and medical \citep{bellovin2019privacy, garfinkel2015deidentification, giuffre2023harnessing}. Traditional evaluations have focused on re-identification risks, particularly for tabular data in census or medical settings \citep{abowd20232010, el2011systematic}. 
Although there have been attempts to create text anonymization benchmarks \citep{pilan2022text}, the benchmarks primarily address span detection and anonymization and focus on scrubbing methods.
Our work examines re-identification and also incorporates modern sanitization methods that generate new data. 
Additionally, researchers have proposed membership inference attacks to determine whether specific data was used in training and  canary attacks to detect the memorization of specific sequences in LMs \citep{carlini2019secret,carlini2021extracting,carlini2022quantifying}.
These privacy evaluation approaches differ from our work because they examine models trained on the data rather than on the dataset itself. 
They depend on access to or assumptions about the model that processed the original text, which is not applicable to all data sanitization techniques.
Recent work in the security literature has begun to challenge the perceived safety of synthetic data \citep{stadler2022synthetic,yale2019assessing,annamalai2024linear}, raising concerns about its privacy guarantees. However, these investigations primarily focus on simple, low-dimensional tabular or image data and have not extended to unstructured text, leaving a critical gap. 

\noindent\textbf{Data sanitization through removal of identifiers.}
To effectively sanitize unstructured text and prevent re-identification, 
researchers have proposed theoretical privacy frameworks such as adapted k-anonymity \citep{venkatesan2008efficient,lison-etal-2021-anonymisation} and mutual-information-based methods \citep{6040853}. 
However, implementing these frameworks is often costly and impractical: text can be expressed in many semantically equivalent forms, making it difficult to define consistent quasi-identifiers for k-anonymity or to reliably measure mutual information across varied expressions of the same content.
These challenges have led to practical data sanitization approaches that focus on detecting and removing Personally Identifiable Information (PII)~\citep{MsPresidio,ines_montani_2022_6397450} using named entity recognition (NER) followed by masking.
Recent work has also explored using LLMs for this task.
\citet{staab2024large} developed an iterative prompting method using GPT-4 to achieve implicit attribute removal, moving beyond simple token replacement. 
\citet{dou2023reducing} proposed a two-step approach, combining a self-disclosure detection model with an abstraction technique to reduce privacy risks in text data.
Other sanitization methods identify sensitive words by
prompting an LLM \citep{zhou2024rescribersmallerllmpowereduserleddata}.
\citet{morrisUnsupervisedTextDeidentification2022} introduced an unsupervised de-identification method that removes words that could lead to re-identification.
However, their method requires a dataset of aligned text and profiles for training, posing unrealistic constraints.
These approaches mainly sanitize the dataset by abstracting or removing detected keywords to minimize re-identification and are susceptible to our proposed semantic re-identification attack.

\noindent\textbf{Data sanitization through synthesis.}
To provide untargeted protection at the dataset level, researchers have employed data synthesis \citep{garfinkel2015deidentification}, occasionally with the assumption that synthesis alone provides some degree of privacy \citep{liu2024best}. 
For a more principled way of providing formal privacy guarantees, differentially private (DP) data synthesis techniques have been developed, including differentially private generative adversarial networks for tabular data synthesis~\citep{xie2018differentially, Torkzadehmahani2019DPCGANDP}.
For textual data, prior work proposed and benchmarked differentially private VAE, BART, and autoencoders with embedding rewards~\citep{weggenmann_dp-vae_2022, igamberdiev_dp-bart_2023,bo_er-ae_2021,Igamberdiev.2022.COLING}, and  \citet{yue2023synthetic,mattern_differentially_2022,mireshghallah2022privacy,kurakin2023harnessing} introduced differentially private fine-tuning approaches for LLMs to generate synthetic text.
More recent works, such as \citet{pang2024reconstructiondifferentiallyprivatetext} and \citet{morris2024diriadversarialpatientreidentification}, show that DP-sanitized records 
can still be linked to the original records, but we further show that DP methods significantly degrade 
their utility. \citet{ramesh2024evaluatingdifferentiallyprivatesynthetic} explored the tradeoff between privacy and utility as well as fairness issues of DP methods on simple classification tasks, using canary evaluation and PII detection to evaluate privacy preservation.
In contrast, our work provides a general, method- and task-agnostic framework for evaluating \textit{semantic} privacy under utility constraints. %

\section{Re-identification Attack Framework} \label{sec:priv-framework}

\paragraph{Threat model.}
Our framework evaluates \textit{re-identification privacy risks in sanitized datasets when released to third parties} for research or collaboration. 
We consider an adversary who: (1) has access to the released sanitized dataset, and (2) possesses auxiliary information about specific individuals from external sources such as social media, public records, or personal acquaintance. 
The adversary's goal is to link this auxiliary information to records in the sanitized dataset to infer sensitive attributes about individuals.

Our threat model differs from existing privacy evaluations that addresses what can be learned about training data from \textit{trained models} (e.g., including membership inference attacks, attribute inference, and training data extraction~\citep{carlini2019secret,stadler2022synthetic,carlini2021extracting,huang2023privacy}.
Instead, we evaluate privacy risks in the \textit{sanitized dataset itself} prior to any downstream use.

\paragraph{Problem statement.}
Let $\Do = \{x^{(i)}\}_{i=1}^N$ denote the original dataset and $\Ds = S(\Do) = \{y^{(j)}\}_{j=1}^M$ the sanitized dataset\footnote{In this paper, we set $N=M$, i.e., the sanitized dataset to be the same as the original dataset. However, this need not be the case for data synthesis methods, which do not require such a one-to-one correspondence.}
for the given data sanitization method $S$. 

Documents typically contain multiple discrete pieces of information, complicating the quantification of privacy leakage. 
For example, Alice's record in Figure~\ref{fig:overview} encompasses both her habits and medical information, making it difficult to assign a single privacy metric that accounts for all sensitive data concurrently.
To address this issue and facilitate a more fine-grained approach to privacy evaluation, we adopt the decompose-then-verify paradigm from FActScore~\citep{minFActScoreFinegrainedAtomic2023b} to atomize each document $x^{(i)}$ into atomic claims ${x^{(i)}_{k}}$---where each claim represents a single, indivisible piece of information---using LLaMA 3.1 8B \citep{dubey_llama_2024}.

Our goal is to evaluate the privacy of $\Ds$ under a re-identification attack by an adversary that has access to $\Ds$ as well as auxiliary information $\tilde{x}^{(i)} = A(x^{(i)})$ for entries in $\Do$. %
The access function $A$, which determines the amount and type of auxiliary information,  depends on the threat model; in our experiments, we set $A(x)$ to randomly select three claims from $x$,
and we maintain the same set of claims across our experiments to ensure consistent comparison.
Furthermore, the access function $A$ paraphrases each claim in the auxiliary information to model potential lexical perturbation during information transmission.
We justify the choice of three claims and examine various designs of the access function by ablating the types of claims used (\S\ref{sec:abl-aux-type}), the effects of paraphrasing (\S\ref{app:ab_para_aux}), and the number of claims (\S\ref{sec:num-claims}).

To assess potential privacy breaches that could result from the public release of a sanitized dataset, we define $L(\tilde{x}^{(i)}, \Ds) \to \hat{y}^{(i)}$ as a linking method that takes some auxiliary information $\tilde{x}^{(i)}$ and the sanitized dataset $\Ds$ as inputs and produces a linked record $\hat{y}^{(i)} \in \Ds$ as output. 
In addition, let  $\mu(x^{(i)}, \hat{y}^{(i)})$ be a semantic distance metric quantifying the dissimilarity between the original record $x^{(i)}$ and the linked record $\hat{y}^{(i)}$.
Given these components, we define our privacy metric as:
\begin{equation}
    \begin{split}
&\text{privacy}(\Do, \Ds) \\= &\mathbb{E}_{x^{(i)} \in\Do}[\mu(x^{(i)}, L(\tilde{x}^{(i)}, \Ds))], 
    \end{split}
\end{equation}
where $\tilde{x}^{(i)} = A(x^{(i)})$ is the auxiliary information as defined above.

In addition, we measure the utility of $\Ds$ to explore the privacy-utility tradeoff, which we detail in \S\ref{sec:utility}.

\paragraph{Linking method $L$.} \label{sec:linker}

We employ a state-of-the-art dense information retrieval technique $L_{\text{dense}}$, specifically, the GRIT retriever \citep{muennighoff2024generative}, to link auxiliary information with sanitized documents. This retriever embeds both the query and the documents into a high-dimensional vector space, enabling semantic similarity. 
For each claim in the auxiliary information ${\tilde{x}^{(i)}}_k \in {\tilde{x}^{(i)}}$,
we perform an independent similarity search against a dataset of atomized sanitized documents.
To select the final document, we use majority voting on each matched claim %
and choose the sanitized document with the highest number of matches across the set of auxiliary information.
We present an ablation study on linker designs in \S\ref{app:ablation-linker}.

\paragraph{Semantic distance metric $\mu$.} \label{sec:fact-priv-metric}

Upon linking auxiliary information $\tilde{x}^{(i)}$ to a sanitized document $\hat{y}^{(i)}$, we quantify the amount of information gain using a semantic distance metric $\mu_{\text{semantic}}$. 
This metric uses an LM to assess the semantic dissimilarity between each claim retrieved from sanitized document $\hat{y}^{(i)}$ and each claim in its original counterpart $x^{(i)}$.
We employ a three-point scale for this assessment: a score of 1 indicates identical information, while 3 signifies that the claim is unsupported by the sanitized document.
We normalize the reported scores to the range [0,1]. 
In this scoring scheme, a higher value of $\mu$ corresponds to a greater degree of privacy preservation. %
Appendix~\ref{sec:privacy-prompt} provides the complete prompt used for this evaluation.

Our implementation uses LLaMA 3.1 8B \citep{dubey_llama_2024} to calculate the semantic distance metric  $\mu$.
To improve the model's consistency, we query the LLaMA model five times for each semantic distance metric evaluation and determine the final classification based on the mode of these responses.
We present a human study of our implementation in 
\S\ref{app:human}.

\paragraph{Lexical baseline.} \label{sec:baseline-metric}
To validate our approach of using semantic information to evaluate privacy preservation, we implement a lexical baseline with functions $L_{\text{rouge}}$ and $\mu_{\text{rouge}}$ implemented using ROUGE-L \citep{lin2004rouge}, widely used in the literature as an automated metric 
\citep{dou2023reducing,xiao2024large,frikha2024incognitext,huang2023privacy}. 
Specifically, the baseline linking method $L_{\text{rouge}}$ processes auxiliary information $\tilde{x}^{(i)}$ by concatenating it into a single text chunk  
and identifies the sanitized document with the maximum ROUGE-L score. To compute the baseline privacy metric $\mu_{\text{rouge}}$, we calculate one minus the ROUGE-L score between the original document $x^{(i)}$ and its linked sanitized version $\hat{y}^{(i)}$. 
This formulation ensures that higher values indicate stronger privacy protection. 
We further explore alternative lexical metric designs by varying the linker in 
\S\ref{app:improving-lexical}, 
different models in \S\ref{sec:model-ablation}, 
and different baseline metrics in Appendix~\ref{app:alt-metrics}.

\section{Experiment Setup}
\subsection{Datasets}

\paragraph{MedQA dataset.}
The MedQA dataset \citep{jin2021disease}  consists of multiple-choice questions derived from the United States Medical Licensing Examination, encompassing a broad spectrum of general medical knowledge. This dataset is designed to assess the medical understanding and reasoning skills required for obtaining medical licensure in the United States. It contains 11,450 questions in the training set and 1,273 in the test set. Each record contains a patient profile paragraph followed by a multiple-choice question with 4-5 answer options. 

\paragraph{WildChat dataset.} \label{sec:wildchat}
The WildChat dataset \citep{zhao2024wildchat} consists of 1 million real user-ChatGPT interactions containing sensitive personal information \citep{mireshghallah2024trust}. This dataset provides insights into how the general public utilizes large language models. Following the pre-processing steps outlined in \citet{mireshghallah2024trust}, we categorize each conversation $x^{(i)} \in \Do$ and task the sanitization method $S$ to generate sanitized conversations. 

In the user-bot interactions, the chatbot frequently produces extensive and repetitive content, particularly when responding to user questions. This behavior reduces the proportion of user-supplied information in the atomization process. To address this issue,
we implement an additional pre-processing step: summarizing each conversation before atomizing the dataset. This prevents the atomization process from being dominated by lengthy chatbot responses.

\subsection{Utility Metrics} %
\label{sec:utility}
\paragraph{Downstream task metrics. } 
Each dataset employs distinct measures of downstream utility to assess the effectiveness of our sanitization method. 

For the MedQA dataset, we evaluate the effectiveness of the sanitized documents in multiple-choice questions. 
We treat the patient profiles as private information requiring sanitization. 
Given a sanitization method $S$, and for each record $x^{(i)} \in \Do$, we generate a sanitized version of the patient profile, and task the evaluation model, LLaMA 3.1 8B \citep{dubey_llama_2024}, with the multiple-choice question using the sanitized patient profile. We report the accuracy of this evaluator's performance as the utility metric.

For the WildChat dataset, we measure the sanitizer's ability to preserve the original conversation distribution since aggressive sanitization can distort records to the point where they are classified into different categories.
We evaluate this by comparing the distribution of categories in generated conversations against the original data, reporting the chi-squared distance as our utility metric.
Following \citet{mireshghallah2024trust}, we use GPT-4o\footnote{\url{https://openai.com/index/hello-gpt-4o/}} as the evaluation model for determining the category. 
We normalize the chi-squared distance on a scale where 1 represents perfect distribution preservation, and 0 corresponds to the chi-squared distance between the original distribution and a uniform distribution across all categories. When a distribution deviates substantially from the original, negative values may occur.

\paragraph{Quality of generation metric. } We furthermore add the sanitization quality metric to our utility metric suite.
Inspired by recent works \citep{zeng_good_2024, chiang_can_2023}, we employ a large language model (in our case, GPT-4o)
as a judge to assess the quality of sanitization outputs on a Likert scale of 1 to 5, with a specific focus on text coherence.

\subsection{Data Sanitization Techniques} \label{sec:san_techniques} %
\begin{figure*}[ht]  
	\centering
	\includegraphics[width=0.7\linewidth]{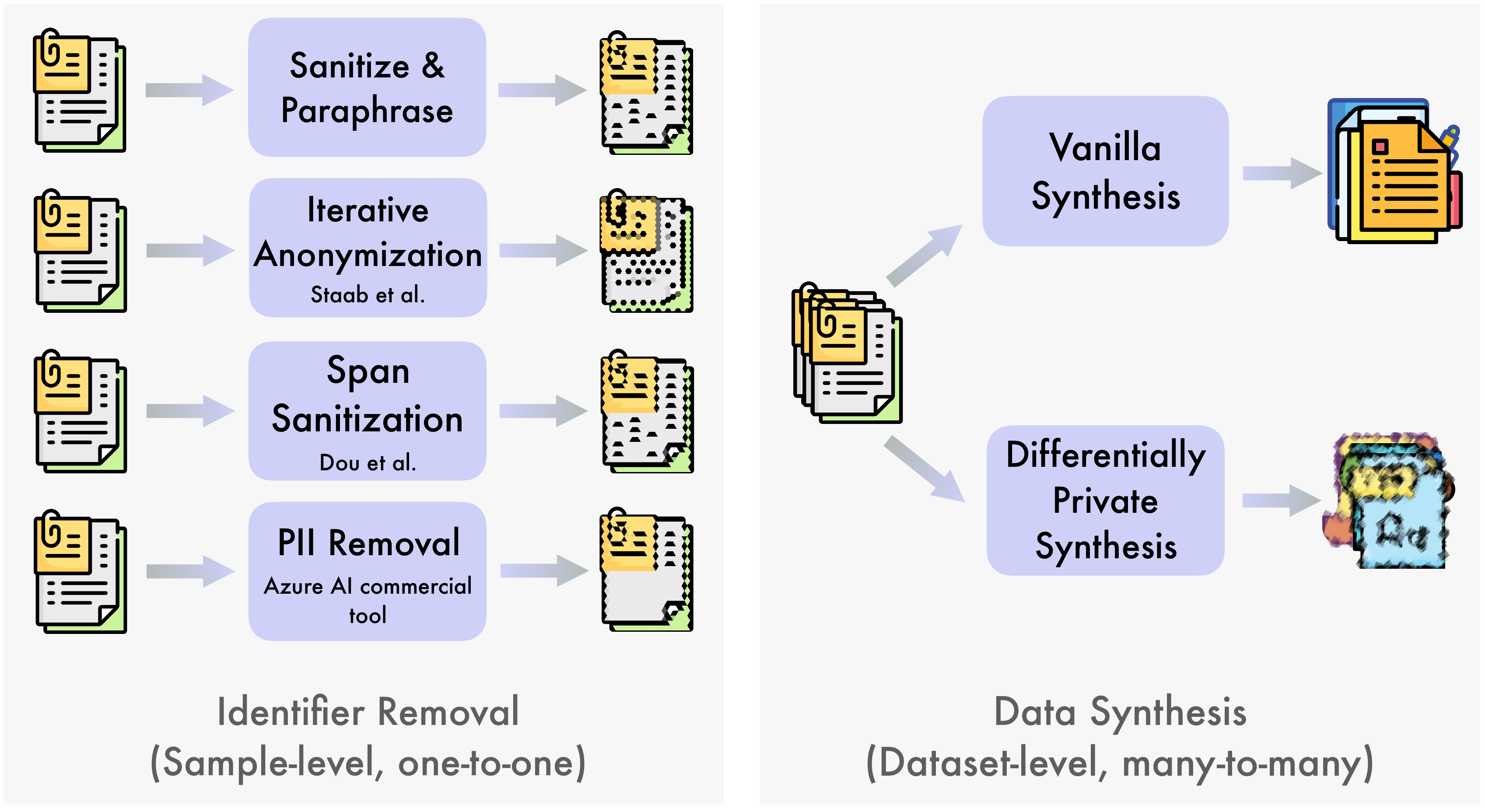}
	\caption{Overview of the data sanitization techniques evaluated using our framework. We evaluate two main categories: identifier removal methods and data synthesis methods. Identifier removal methods operate at the sample level, maintaining a one-to-one correspondence between original and sanitized records. In contrast, data synthesis methods operate at the dataset level, where each sanitized record may derive information from multiple original records.}

	\label{fig:experiments}
\end{figure*}

We use our metrics to evaluate two categories of sanitization methods, as illustrated in Figure~\ref{fig:experiments}.
Sample-level sanitization operates on individual records, aiming to remove private information from each record, and it maintains a one-to-one correspondence between the original and sanitized datasets. In contrast, dataset-level sanitization seeks to create a new dataset that preserves the the textual patterns and linguistic characteristics of the input dataset, 
where sanitized records may not directly correspond to those in the original dataset. Detailed prompts used in our analysis are provided in Appendix~\ref{sec:prompts}.

\paragraph{Iterative anonymization \citep{staab2024large}.} \label{sec:sanviaprompt}
This approach utilizes LLMs 
to remove sensitive information through iterative prompting. We implement the sanitization pipeline proposed by \citet{staab2024large}, which employs a two-step process of adversarial inference and sanitization. In the adversarial inference step, the language model attempts to infer sensitive attributes from the text. Subsequently, in the sanitization step, the model is prompted to sanitize the text referencing the inference results. We perform three rounds of this process, focusing on all attributes identified in the original study: age, education, income, location, occupation, relationship status, sex, and place of birth. For this sanitization method, we employ GPT-4o as our LLM.

\paragraph{Sanitize and paraphrase.} \label{sec:sanviaprompt_paraphrase}
Drawing insights from \citet{zeng_mitigating_2024}, who explored record rewriting, we implement a sequential privacy protection approach.
We first apply the sanitization prompt from \citet{staab2024large} without attribute inference, then use GPT-4o to paraphrase the sanitized text, potentially enhancing privacy protection.

\paragraph{Span sanitization \citep{dou2023reducing}.}
We evaluate the self-disclosure detection model developed by \citet{dou2023reducing}. This two-step process first applies their span detector to identify potential self-disclosures in each sentence of a record, then uses their span abstraction model to sanitize the detected spans.

\paragraph{Azure AI PII tool.} \label{sec:san_via_scrubbing}
We evaluate an industry grade data sanitization method that focuses on identifying and removing personally identifiable information (PII). This approach utilizes the Azure AI Language PII detection service\footnote{\url{https://learn.microsoft.com/en-us/azure/ai-services/language-service/personally-identifiable-information/overview}} to identify and redact PII from the dataset with the ``*'' character.

\paragraph{Data synthesis via differentially private fine-tuning.}
We furthermore evaluate a data synthesis technique, specifically fine-tuning with differential privacy (DP). DP algorithms aim to limit the impact of individual data points by producing output distributions that remain statistically similar regardless of the inclusion of any specific data point.
We adopt the method described by \citet{yue2023synthetic}, which generates synthetic text while maintaining formal DP guarantees. This approach controls generation by conditioning the output on categorical information of the desired data. Prior to fine-tuning a generative model, the method preprocesses data records by prepending a 
``control code'', a categorical label, to each record. 
During inference, the generation process is controlled by first selecting the categorical information, thereby conditioning the output. We discuss the scope and limitations of this specific DP method in Appendix~\ref{sec:limitations}.

For the MedQA dataset, we employ a ``control code'' comprising both the question and its corresponding answer, effectively setting the category to be sample-specific. Specifically, we prepend a text snippet in the format ``Question: [question text] \textbar Answer: [answer text]'' to each record $x^{(i)}$. During the generation of sanitized records, we provide this same text snippet and ask the model to generate the corresponding record,
treating the generated record as the sanitized information. 
We treat the question and answer as \textit{public information} and the patient profile as \textit{private information}.

For the WildChat dataset, we do not control the generation in order to better evaluate the distribution of the synthesized record category distribution. 

In our experiments, we apply this method to our datasets with privacy budget values of $\varepsilon \in \{3, 8, 16, 64, 512, 1024\}$ that are commonly used in the differential privacy literature; see Appendix~\ref{app:dp-design} for detailed justification of the design choice.

\paragraph{Data synthesis via language model fine-tuning.}
We implement a data processing pipeline following the approach described above. The implementation uses the same control code mechanism but with standard fine-tuning parameters: an unbounded privacy budget ($\epsilon = \infty$) achieved by disabling noise injection and gradient clipping.

\paragraph{Sanitization baselines.}
We incorporate two additional baselines: \textbf{No Sanitization} and \textbf{No Private Data}. The \textbf{No Sanitization} baseline utilizes the original, unmodified text to establish a performance reference point, serving as both a lower bound for privacy protection and an upper bound for data utility. Conversely, the \textbf{No Private Data} baseline, evaluated on MedQA, sanitize the text by removing the provided record, which is considered as private, measuring the underlying knowledge and inherent biases of the language model.

\section{Experiment Results}

\subsection{The Privacy-Utility Tradeoff: Identifier Removal and Data Synthesis} \label{sec:main-result}

\begin{table*}[h!]
    \centering
    \begin{minipage}{\textwidth}

\input{tables/sanitization_table_comp}

    \end{minipage}

    \begin{minipage}{\textwidth}

\input{tables/sanitization_table_dp}

    \end{minipage}
\end{table*}

Table~\ref{tab:sanitization-comparison} shows our results on the privacy-utility tradeoff of various data sanitization techniques. 
Our evaluation uses two privacy metrics, i.e.,  semantic distance and lexical distance (our baseline approach from \S\ref{sec:priv-framework}) and two utility metrics, i.e., task utility and text coherence (discussed in \S\ref{sec:utility}).

Identifier removal methods (Sanitize \& Paraphrase, Azure AI PII tool, Span Sanitization, Iterative Anonymization) display a consistent pattern: their lexical distance values exceed their semantic distance. 
\textit{This difference reveals that while these methods alter surface text, they preserve the underlying semantic connections that enable inference attacks. }
The Azure AI tool, despite its commercial adoption, achieves only 0.26 semantic distance, indicating that 74\% of its information is still available after sanitization.
This failure occurs because commercial PII tools focus exclusively on removing explicit identifiers (names, addresses, dates) while ignoring semantic patterns that enable re-identification. As shown in Figure~\ref{fig:leakage_sensitive_category}, Azure AI successfully removes direct identifiers but fails to sanitize semantic categories such as ``History of Present Illness,'' ``Social History,'' and ``Diagnostic Results''---categories that contain rich contextual information enabling inference attacks. Table~\ref{tab:examples} in the appendix provides concrete examples of how our attack successfully re-identifies records and infers sensitive attributes like age, medical conditions, and substance use history from sanitized data.
Among these data sanitization methods, Iterative Anonymization \citep{staab2024large} outperforms other identifier removal methods, but it sanitizes only a predefined set of attributes, resulting in reduced performance when claims fall outside the established category list. 
Similarly, Span Sanitization \citep{dou2023reducing} requires a list of categories to detect and sanitize.

Data synthesis methods reduce the gap between lexical and semantic privacy metrics compared to identifier removal approaches; thus, \textit{they produce less surface-level perturbation while still altering the semantic meaning of the underlying text.}
Their effectiveness varies by dataset. 

On MedQA, data synthesis methods achieve privacy and utility levels similar to identifier removal.
On WildChat, data synthesis results in higher privacy at the expense of lower task utility compared to most identifier removal methods.
This difference stems from the varied approaches to data synthesis across datasets. For the MedQA dataset, we conditioned the LM on both questions and answers to generate patient profiles that maintain utility.
For the WildChat dataset, we did not control the generation since we focused on preserving the distribution of conversation categories. This approach increased the semantic distance, indicating that the sanitized documents differ substantially from the original ones, resulting in higher privacy scores.

These results demonstrate that
lexical metrics, when applied to identifier removal methods,
create a \textit{false sense of privacy}. 
Lexical metrics report artificially higher privacy than actual semantic information leakage. 
In contrast, lexical metrics align more closely with semantic metrics when evaluating data synthesis methods, particularly for the WildChat dataset.
However, this better representation of privacy value corresponds to a scenario where the utility degrades.

We further explore an alternative lexical metric design that improves on common techniques in the literature (Appendix \ref{app:improving-lexical}). By replacing the semantic linker with a sparse linker, we observed that the lexical-semantic gap decreases while lexical privacy continues to exceed semantic privacy for most of the sanitization methods. This suggests that \textit{lexical scores continue to overstate actual privacy protection.}

\subsection{The Privacy-Utility Tradeoff: Data Synthesis with Differential Privacy}

In the previous section, we showed that data synthesis offers a privacy-utility tradeoff similar to identifier removal methods. 
However, this sanitization technique remains imperfect since privacy leakage persists. 
Table~\ref{tab:sanitization-comparison-dp} evaluates the previously discussed metrics under differential privacy (DP) guarantees.
Researchers often integrate data synthesis with DP to establish formal bounds on potential data leakage \citep{yue2023synthetic}.
Bounding the leakage in DP is governed by the privacy budget, denoted as $\varepsilon$. A higher $\varepsilon$ value corresponds to reduced privacy. 
The row where $\varepsilon=\infty$ is equivalent to not applying differential privacy, i.e., the 
data synthesis row from Table~\ref{tab:sanitization-comparison}.

\textit{Results show that applying DP improves privacy protection even with high privacy budgets} such as  $\varepsilon=1024$. For MedQA, the lexical privacy metric increases from $0.76$ to $0.84$, and the semantic privacy metric from $0.52$ to $0.90$. These privacy improvements come at the cost of utility. 
The MedQA utility decreases from $0.61$ to $0.42$, dropping below the $0.44$ no private data baseline; we discuss this counterintuitive result in Appendix~\ref{app:dp-design}.

The WildChat dataset exhibits similar utility degradation under DP. With a strict privacy budget ($\varepsilon=3$), the utility falls below $0$, indicating that the sanitized label distribution deviates from ground truth more than a uniform distribution would. The textual coherence metric also decreases substantially from $3.28$ to $1.86$, where $1$ represents ``Very Poor'' quality text. We show an example output in Table~\ref{tab:eps3_example}.

\begin{table}[h]
\caption{A medical record generated by the DP sanitization method with $\varepsilon=3$. 
We note that the record suffers from semantic inconsistencies, including contradictory statements about the patient's health status and redundant physical examination mentions. These artifacts are typical of DP-generated text, where coherence is compromised to maintain privacy guarantees.}
\centering
\vspace{1ex}
\begin{tabular}{p{0.95\columnwidth}}
\toprule
A Sample Medical Record Generated by the DP Sanitization Method with $\varepsilon=3$: \\
\midrule
A 21-year-old man presents to his family physician for evaluation\ldots 
On physical examination, he is in good general health and
\textbf{his physical examination reveals no abnormalities}. His pulse is 116/min.
His temperature is 37.7°C (100.4°F), blood pressure is 103/73 mm Hg,
and body weight is 62 kg (139 lb).
\textbf{Physical examination shows generalized tenderness throughout
the back and extremities}, along with an intermittent, 
tender warmth on the neck and forehead \ldots \textbf{Examination of his abdomen reveals a 4-mm-long papillary mass} \ldots \\
\bottomrule
\end{tabular}
\label{tab:eps3_example}
\end{table}

Unlike non-DP results, some $\varepsilon$ settings produce lexical privacy metrics that are lower than semantic similarity metrics. Through manual inspection, we found that this occurs due to the degraded text quality. These cases show minimal meaningful information leakage, with non-perfect lexical privacy scores ($<1.0$) arising from matches in common words like articles and prepositions rather than from actual private content leakage.

\section{Analysis}
\subsection{Changing the Available Auxiliary Information} \label{sec:abl-aux-type}

\input{tables/aux_info_table}

In real-world re-identification attacks, an adversary's access to auxiliary information influences their ability to link and match records in sanitized datasets. 
For example, in the MedQA dataset, the first three claims often contain a fixed set of information, such as the age, sex, and chief complaint of the patient, while the last three claims lack such  information and are filled with arbitrary facts such as lab results. 
Our previous experiments randomly selected three claims from each record as the adversary's accessible information. To assess the impact of this choice, we conducted experiments using (1) randomly selected claims, (2) the first three claims, as well as (3) the last three claims. 

Table~\ref{tab:sanitization-aux} presents the results of these experiments, focusing on the \textit{correct linkage rate}. This metric 
quantifies the percentage of correctly paired original and sanitized documents when using the provided auxiliary information in cases where ground truth relationships are known.

Results demonstrate that \textit{the type of auxiliary information available to the adversary affects the linking step}, providing insights into the effectiveness of various sanitization methods.
For the MedQA dataset, we observe
that the methods relying on LLMs, such as Sanitize \& Paraphrase and the approach proposed by \citet{staab2024large} show the largest linkage differences between the first three and last three claims. In contrast, No Sanitization reveals minimal differences in linker performance based on claim type, indicating that the observed differences are specific to the sanitization methods rather than inherent to the data.
We hypothesize that this variation in linkage rates occurs because LLMs are more effective at sanitizing certain information types, such as patient age, that appear more frequently in earlier claims, resulting in uneven information preservation across different sections of the text.

\subsection{Ablating Treatment on the Auxiliary Information} \label{app:ab_para_aux}
\input{tables/linker_table_paraphrase_aux}

We examine how perturbing auxiliary information affects our privacy metric, which we use to simulate lexical changes that occur to auxiliary information during transmission, as described in \S\ref{sec:priv-framework}. 
We implement this perturbation because obtaining original data, such as protected medical records or exact conversation transcripts, is rarely possible in practical scenarios.
For example, we paraphrase the original auxiliary information ``Auscultation of the lungs does not reveal any significant abnormalities.'' into ``A thorough examination of the patient's lungs did not uncover any notable issues.'' 
Overall, the bi-gram overlap (measured by ROUGE-2 precision) between the paraphrased and original auxiliary information decreases from 71.0\% to 13.0\% for MedQA and from 40.5\% to 17.7\% for WildChat. %

To evaluate the impact of this design choice, we conduct our privacy analysis using both the original and paraphrased auxiliary information, with results shown in Table~\ref{tab:ablation-aux-info}.
The relative effectiveness of different sanitization methods remains consistent across both conditions---methods with higher leakage using original auxiliary data also show higher leakage with paraphrased data.
This property is essential for privacy evaluation in practical situations where exact replicas of sensitive information are unavailable.

\subsection{Ablation on Linker} \label{app:ablation-linker}
We conduct experiments on different linker designs, focusing on two key aspects: comparing retrieval methods and evaluating strategies to construct retriever queries with the auxiliary information. Our analysis contrasts GRIT, a semantic retriever, with BM25 \citep{Lin_etal_SIGIR2021_Pyserini}, a sparse retriever, while also examining different approaches to construct the query for the retriever. 
In this ablation study, we evaluate the effectiveness of various linker designs using the correct linkage rate metric. 
This metric quantifies the percentage of original and sanitized document pairs that are correctly matched when using the provided auxiliary information.

\paragraph{Varying retriever.}
Our baseline implementation uses GRIT \citep{muennighoff2024generative}, a dense retriever that matches auxiliary information to sanitized documents that embeds both queries and documents in a high-dimensional vector space and retrieves nearest neighbors based on semantic similarity. We compare this against BM25 \citep{Lin_etal_SIGIR2021_Pyserini}, using term frequency-inverse document frequency (TF-IDF) weighting. 

\paragraph{Varying query construction from auxiliary information.}
We evaluate two approaches to construct queries from auxiliary information. 
Our primary method treats each piece of auxiliary information ${\tilde{x}^{(i)}}$ as an independent query against the database of atomized sanitized documents.
The final document selection uses majority voting, selecting the document that most frequently matches across all auxiliary information claims.
In the second approach, we merge all auxiliary information into a single query. 
This design allows the retriever to observe all information simultaneously, but results in decreased granularity.

\input{tables/linker_table}

Results in Table~\ref{tab:linker} show that the GRIT retriever with majority voting performs better than other linkers on most sanitization methods on the MedQA dataset. 
This effectiveness stems from our approach to paraphrase auxiliary information before we feed it into the retriever as the query, as a sparse retriever is unable to perform well when the exact phrases have been changed.
The merged single query perform worse on many sanitization methods. 
We attribute this to the semantic retriever's improved performance when matching single pieces of information, particularly in datasets like MedQA where each record contains information pieces with minimal overlap.

The WildChat dataset exhibits a similar pattern. As a dataset of user-chatbot interactions, it contains a more diverse and common vocabulary compared to the specialized medical terminology in MedQA. 
The merged query approach performs at least as well as majority voting across most sanitization techniques, except for the no-sanitization condition.
We attribute this to WildChat documents typically containing unified themes, where comprehensive information provides better context for matching.
This contrasts with MedQA, where individual pieces of auxiliary information have fewer overlaps.

\subsection{Improving Lexical Metric} \label{app:improving-lexical} 
In \S\ref{sec:priv-framework}, we implement the lexical metric baseline using a standard lexical metric, ROUGE \citep{lin2004rouge}. 
We employ it to serve as both the linker $L_{\text{rouge}}$ and the method for computing the final metric $\mu_{\text{rouge}}$, following common practice in current literature.
However, this approach is limited when processing paraphrased auxiliary information, as ROUGE scores decrease with changes in sentence structure or vocabulary, even when semantic meaning is preserved.
To address this limitation, we enhance the linker by implementing a sparse retrieval method, BM25, as introduced in \S\ref{app:ablation-linker}. BM25 offers improved matching capability for paraphrased content by focusing on term frequency and inverse document frequency rather than exact sequence matching. This approach emphasizes rare words that often persist through paraphrasing, enabling more accurate document linkage.

\input{tables/improve_lexical}

Table \ref{tab:improve-lexical} shows the results when applying this improved lexical metric, that we label as \textbf{BM25 Matching + Rouge Metric}. While the gap between lexical and semantic distance decreases, lexical privacy values mostly exceed semantic privacy values, confirming that lexical scores overestimate the actual privacy protection. Additionally, we observe that the Sanitize \& Paraphrase methods show the largest difference between the improved lexical distance and the semantic distance. This indicates that lexical metrics remain inadequate for accurately evaluating privacy in cases where sanitization methods substantially rewrite content, whereas semantic privacy metrics can capture these changes more effectively.

\subsection{Robustness Across Different Models} \label{sec:model-ablation}

To measure the robustness of our privacy evaluation framework across different language models, we conduct additional experiments using GPT-4o on the MedQA benchmark with various sanitization methods. We randomly select new auxiliary information for attackers
since the list of atoms varies across runs. We perform one run for each GPT-4o experiment.

\begin{table}[h]
\centering\vspace{-2mm}
\caption{Privacy scores across sanitization methods comparing GPT-4o and Llama 3.1 8B models on MedQA.}
\vspace{1ex}
\begin{tabular}{lll}
\toprule
\textbf{Sanitization Methods} & \textbf{GPT-4o} & \textbf{\makecell[l]{Llama 3.1\\ 8B}} \\
\midrule
Sanitize \& Paraphrase & 0.25 & 0.33 \\
Azure AI PII tool & 0.18 & 0.26 \\
Span Sanitization & 0.59 & 0.55 \\
Iterative Anonymization & 0.55 & 0.53 \\
Data Synthesis & 0.52 & 0.52 \\
\makecell[l]{Differentially Private Fine-tuning\\ with \(\epsilon=1024\)} & 0.98 &
0.90 \\
\bottomrule
\end{tabular}
\label{tab:model-ablation}
\end{table}

Table \ref{tab:model-ablation} shows the results. We find that the
privacy evaluation trend is consistent with evaluating using the Llama
3.1 8B models, and in fact, further substantiate our claim that it is
essential to move beyond lexical privacy metrics.

GPT-4o privacy metrics show similar trends compared to Llama 3.1 8B. For
low-privacy sanitization methods such as No Sanitization, Sanitize \&
Paraphrase, and Azure AI PII tool, GPT-4o reports slightly lower privacy
scores, indicating more severe privacy risks. We attribute this
difference to the more advanced model's superior ability to infer
content from sanitized information. For mid-privacy methods such as Span
Sanitization, Iterative Anonymization, and Data Synthesis, GPT-4o
reports similar performance to Llama. For differential private
fine-tuning with $\epsilon$=1024 (representing the highest privacy leakage among
our differential privacy experiments), GPT-4o reports higher privacy
scores, confirming the effectiveness of these theoretical privacy
guarantees. However, as emphasized in our paper, enhanced privacy comes
at the cost of reduced utility.

The similarity between privacy scores across models echoes our human
study findings. These scores remain substantially lower than lexical
similarity scores for non-differentially private methods, emphasizing
the importance of our semantic-based privacy metric.

\subsection{Human Evaluation of the Semantic Distance Metric} \label{app:human}
To validate our language model's performance in measuring the semantic distance metric $\mu$ defined in \S\ref{sec:fact-priv-metric}, we conducted a controlled human evaluation study. Three authors independently annotated 580 identical claims, working without access to any model-generated outputs to prevent bias. 
The evaluation yielded strong inter-annotator reliability, with a Fleiss' kappa coefficient of 0.87. When comparing model performance to human judgments, we found LLaMA 3 8B achieved a Spearman correlation coefficient of 0.95 with the mode of human annotations. This performance approaches that of GPT-4, which achieved a coefficient of 0.97. For comparison, the ROUGE algorithm showed weaker alignment with human judgments, reaching a Spearman coefficient of 0.81.

\subsection{Disentangling Information Leakage Sources} \label{sec:num-claims}

\begin{figure*}[t]
	\centering
	\includegraphics[width=\linewidth]{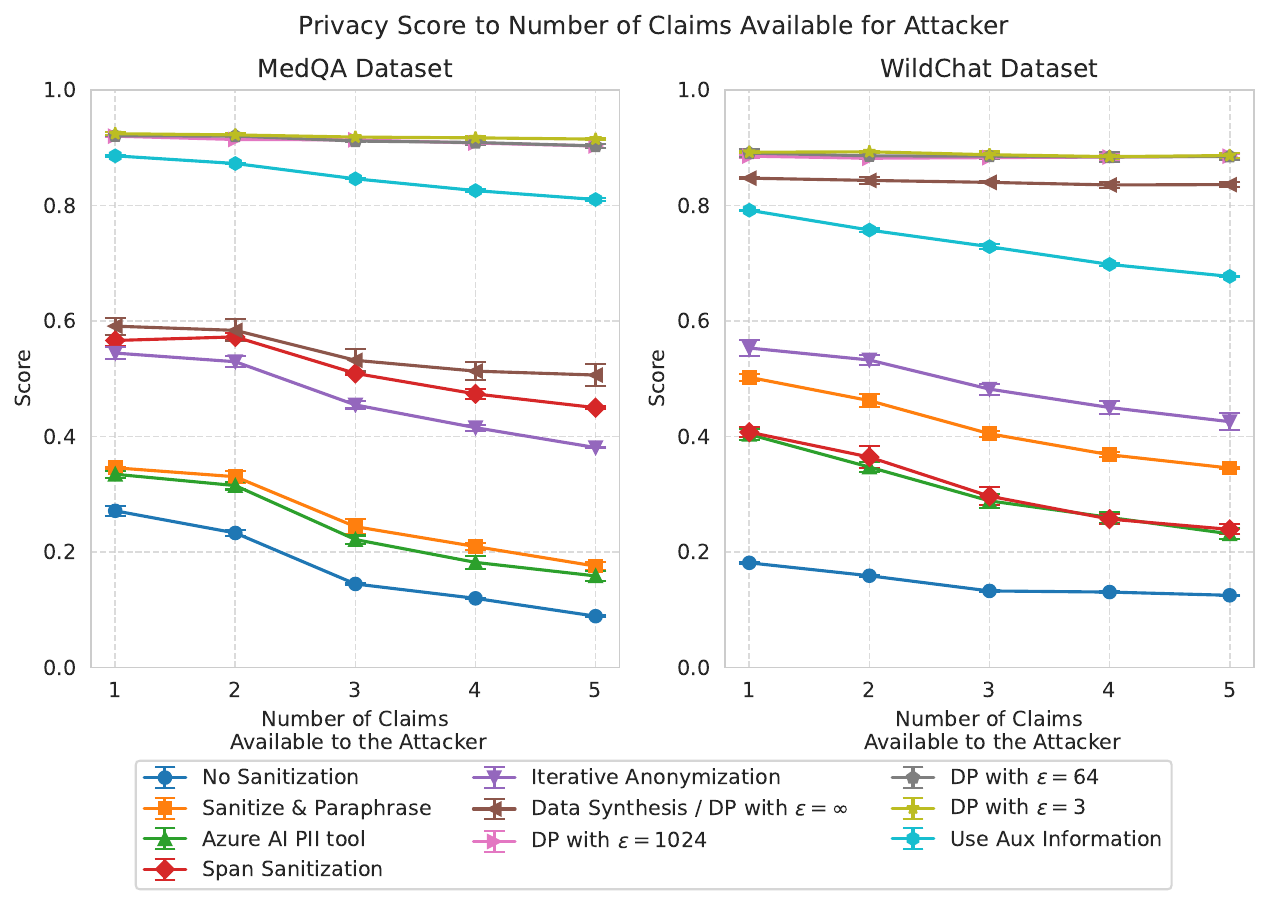}
\caption{Privacy scores to the number of claims available to the attacker across different sanitization methods (\S\ref{sec:san_techniques}). Sanitization methods are introduced in \S\ref{sec:san_techniques}. In particular, \textbf{Span Sanitization} refers to sanitization method proposed by \citep{dou2023reducing}, and \textbf{Iterative Anonymization} refers to the technique proposed by \citep{staab2024large}. The \textbf{Use Aux Information} row quantifies the information overlap between auxiliary information provided to attackers and the remaining document content.}
	\label{fig:ablation_amount_info}
\end{figure*}

We investigate whether detected privacy leakage comes from the sanitized dataset or could be inferred from auxiliary information alone.
We examine how our method's effectiveness varies with both the quantity of information available to the attacker and the information overlap between auxiliary information and the rest of the record.

We first explore privacy score degradation as attackers gain access to more information. Instead of providing three random claims from a record, we provide the last $k$ claims to the attacker, where $k \in \{1, 2, 3, 4, 5\}$, and measure the resulting privacy score.

Then, we investigate the amount of information that can be inferred from auxiliary information alone, without any access to the sanitized dataset. To measure this, we apply the semantic distance metric defined in \S\ref{sec:privacy-prompt}, treating the provided auxiliary information as if it were the sanitized document. In this context, a higher privacy score indicates that less information can be inferred from auxiliary information alone.

Figure~\ref{fig:ablation_amount_info} presents both the privacy metric degradation and the information available from auxiliary data alone, reported as \textbf{Use Aux Information}.
Methods without theoretical guarantees show decreased privacy as attacker information increases, with the steepest decline occurring when adding claims to the case of the last one or two claims. This decline slows with additional claims, supporting our choice to use three claims for sanitization method evaluation. In contrast, DP methods maintain consistent performance regardless of the number of claims available to attackers, demonstrating their robust privacy protection. 

The ``Use Aux Information'' baseline shows modest leakage: auxiliary information alone reveals 20\% of original information for MedQA (privacy score 0.80) and 35\% for WildChat (privacy score 0.65). 
In contrast, when attackers access both auxiliary information and sanitized datasets, non-DP methods leak 45--74\% of information (privacy scores 0.26--0.55). 
This gap demonstrates that detected privacy leakage comes primarily from the sanitized dataset itself, not from inferences based on auxiliary information alone.

\begin{table}[h]
\centering\vspace{-2mm}
\caption{Inter-rater agreement and model correlations for semantic similarity inference task.} %
\vspace{1ex}
\begin{tabular}{llll}
\toprule
\textbf{Metric/Model} & \textbf{Measure} & \textbf{Value} \\
\midrule
Human Agreement & Fleiss' Kappa & 0.875 \\
\midrule
LLaMA 3 8B & Spearman Correlation & 0.919 \\
\midrule
GPT-4o & Spearman Correlation & 0.946  \\
\midrule
ROUGE-L recall & Spearman Correlation & -0.806  \\
\bottomrule
\end{tabular}
\label{tab:agreement-correlations}
\end{table}

\section{Conclusion}
This paper introduces 
a novel semantic-based, dataset-level privacy metric that addresses key limitations in current data sanitization methods for unstructured text. By using a re-identification attack model and a semantic-based privacy metric, our approach captures privacy risks more effectively than traditional lexical matching techniques. 
We emphasize that our attack provides a \textit{lower bound} on information leakage: a low privacy score definitively indicates vulnerability, though a high score does not guarantee safety against all possible attacks. 

Our framework integrates both privacy and utility evaluation for the sanitized dataset, providing a comprehensive evaluation of the tradeoffs involved in different sanitization techniques. 
Experiments on MedQA highlight that although differential privacy provides strong privacy protection, it often dramatically reduces data utility. Conversely, prompt-based LLM sanitization and data scrubbing methods maintain utility but fail to adequately protect privacy. 
Fine-tuning offers privacy-utility tradeoffs  similar to identifier removal methods for the  MedQA dataset but suffers from low utility on the WildChat dataset.

This work advances privacy evaluation by providing a holistic framework that helps  researchers better navigate the tradeoffs between privacy and utility and provides a test bed for future research in data sanitization. 
Our experiments reveal that existing sanitization methods often create \textit{a false sense of privacy} by implementing text modifications at the surface level without addressing deeper semantic vulnerabilities.
Our results highlight the urgent need for new privacy protection methods that specifically target the problem of semantic information leakage while preserving utility. We acknowledge that achieving both strong privacy protection and high utility remains an open challenge---to date, no methods outside differential privacy have offered robust protection against privacy attacks, yet DP methods incur substantial utility costs on complex tasks.

\section*{Acknowledgments}

We would like to thank 
Staab Robin, Yao Dou, Hamish Ivison, Siting Li, Scott Geng 
for insightful discussions. 
This research was supported by the University of Washington Population Health Initiative, as well as the Singapore National Research Foundation and the National AI Group in the Singapore Ministry of Digital Development and Information under the AI Visiting Professorship Programme (award number AIVP-2024-001), the AI2050 program at Schmidt Sciences, Meta AIM program, Coefficient Giving, Amazon Health, NSF awards 2112471, 2229876, 2505865, and 2142739, the UW+Amazon Science Hub, and gift funding from MSR and Google.
This research was developed in part with funding from the Defense Advanced Research Projects Agency's (DARPA) SciFy program (Agreement No. HR00112520300). and the DARPA and the Air Force Research Laboratory, contract number(s): FA8650-23-C-7316. Any opinions, findings and conclusions, or recommendations expressed in this material are those of the author(s) and do not necessarily reflect the views of AFRL or DARPA.

\bibliography{colm2025_conference}

\bibliographystyle{IEEEtranN}

\newpage
\appendix

\lstset{
    breaklines=true,
    basicstyle=\ttfamily
}

\subsection{Limitations} \label{sec:limitations}
Our study is not exhaustive, and particularly it does not encompass all possible privatization techniques, such as model unlearning techniques where it is not readily applicable to the data sanitization setting. 
Additionally, our analysis was primarily confined to datasets within the medical and conversational domains, which limits the generalizability of our findings. Future research should focus on evaluating the method's applicability across diverse datasets and domains to establish its broader relevance and robustness.

\paragraph{Scope of differential privacy evaluation.}
Our DP results reflect the limitations of a specific DP-SGD fine-tuning implementation~\citep{yue2023synthetic}, not fundamental constraints of differential privacy itself. 
More advanced DP text generation techniques may achieve better privacy-utility tradeoffs than those observed in our experiments.
Our framework provides a valuable testbed for evaluating such future advances in DP methods for text.
See Appendix~\ref{app:dp-design} for detailed justification of our DP experimental design choices.

\paragraph{Interpretation of privacy scores.}
Our attack provides a \textit{lower bound} on information leakage rather than comprehensive privacy certification. A low privacy score definitively indicates vulnerability, but a high score does not guarantee safety against all possible attacks---more sophisticated attacks may reveal additional vulnerabilities. Conversely, in rare cases (primarily with DP outputs, as noted in Appendix~\ref{app:score-distribution}), high attack success rates may result from language model hallucinations rather than actual privacy leakage.

\paragraph{Information leakage attribution.}
A key challenge in our work is that the definition of privacy and what constitutes a privacy leak is often blurry and context-dependent. Privacy is fundamentally based on outcomes and how people feel about information disclosure, rather than purely objective measures or monetary harm. Our metric measures semantic similarity, which may be more relevant for some types of information (e.g., medical conditions) but less meaningful for others (e.g., social security numbers). This limitation is particularly relevant when comparing our method to techniques specifically designed for PII removal. Furthermore, there is an inherent ambiguity in distinguishing between information learned from the sanitized dataset and information that can be inferred from the auxiliary data. For example, if the auxiliary data suggests that someone is going through mental status examination, one might infer a high probability of mental disease without accessing the sanitized data. 
We address this concern in \S\ref{sec:num-claims}: the ``Use Aux Information'' baseline shows that auxiliary information alone reveals only 20--35\% of original information, while sanitized datasets leak 45--74\%, demonstrating that detected leakage comes primarily from the sanitized dataset.

Our work does not pass judgment on whether or not inferences from the auxiliary data  are privacy violations as some might be necessary for maintaining downstream utility.
Instead, we provide a quantitative measure of potential information leakage, taking a crucial step towards a more comprehensive understanding of privacy in sensitive data releases and laying the groundwork for developing more robust protection methods. Ideally, a more desirable solution would be a \textit{contextual} privacy metric, which can take into account (i) which information is more privacy-relevant and (ii) which information is private in the context that the textual information is being shared. These are challenging questions that we believe are beyond the scope of this paper. Nevertheless, they represent exciting research directions to pursue, particularly given recent advances in LLMs.

\subsection{Differential Privacy Experimental Design} \label{app:dp-design}

\paragraph{Public vs. private information in MedQA.}
For MedQA, we prepend a text snippet in the format ``Question: [question text] | Answer: [answer text]'' to each record as the control code.
We treat the question and answer options as \textit{public information}, while the patient profile paragraphs constitute the \textit{private information} requiring protection. This reflects a realistic scenario where adversaries possess auxiliary information from public sources.

To quantify potential leakage through control codes, we measured the semantic overlap between questions/answers (control codes) and patient profiles (private data) using the same semantic distance metric from our framework. The semantic distance is 0.87. 
This means that 87\% of patient profile information is not present in the control codes and receives full DP protection.

\paragraph{No one-to-one mapping.}
There is no one-to-one mapping between original and DP synthetic texts. The DP model generates new patient profiles conditioned only on public information (question and answer), not the original private profiles. For utility evaluation, we ask an evaluator model to select the correct answer based on these generated profiles alone, measuring whether synthetic profiles contain medically meaningful information.

\paragraph{Choice of $\varepsilon$ values.}
Our choice of $\varepsilon=3$ as the minimum value follows established practices in the DP text generation literature~\citep{yue2023synthetic}. As shown in Table~\ref{tab:eps3_example}, at $\varepsilon=3$ the textual coherence drops substantially (from 3.48 to 2.04), with generated text containing semantic inconsistencies such as contradictory statements about the patient's health status. This utility degradation informed our decision not to evaluate stricter privacy settings with lower $\varepsilon$ values.

\subsection{Additional Experiments}
\subsubsection{Comparison to Alternative Metrics} \label{app:alt-metrics}

\input{tables/diff_metrics}

We evaluate our metrics against other established approaches in measuring privacy preservation, including distributional, embedding-based, and identifier-based metrics.

\paragraph{MAUVE.} We use MAUVE \citep{pillutla2021mauve} to measure the difference between original and sanitized texts using divergence frontiers. This metric does not utilize auxiliary information linking, and instead directly measuring differences between the original and sanitized datasets.

\paragraph{Embedding.} We use the all-MiniLM-L6-v2 model \citep{wang2020minilm} to compute embedding distances between linked original and sanitized documents. We first embed each claim from both original and sanitized documents. Then, for claims not used for linking in the original document, we compute dot products of the selected claim embedding with all sanitized claims and select the maximum score. The final metric represents the mean score across all original document claims.

\paragraph{PII existence.} This baseline metric examines personally identifiable information (PII) detected by Azure AI, excluding information used for document linking. We calculate the match rate between original and sanitized documents for each PII instance.

\paragraph{Lexical and semantic distance.} We include these metrics from \S\ref{sec:main-result} as reference points for our comparison.

The results are shown in Table~\ref{tab:common_metrics}, revealing limitations in existing metrics. MAUVE is inadequate for privacy preservation measurement. For example, in the MedQA dataset, MAUVE reports that Data Synthesis sanitization leaks all information, and it suggests that the PII sanitization is more private compared to Data Synthesis method, achieving a score of 0.51. However, upon manual inspection, it is clear that PII sanitization leaks more information than Data Synthesis. This discrepancy stems from MAUVE's focus on token distribution at the dataset level, ignoring individual record privacy.
The embedding metric, while operating at the record level, is harder to interpret when compared to our semantic distance metric. The maximum score of 0.65 lacks clear privacy implications. 
PII Existence metrics suggest strong privacy preservation for the PII removal method, particularly in the MedQA dataset. However, our analysis reveals that PII sanitization provides little privacy protection, contrary to what this metric suggest.

\subsubsection{Analysis of Categories of Detected Privacy Leakage } \label{app:category-analysis}
\input{tables/category_table}

\begin{figure*}[]
	\centering
	\includegraphics[width=\linewidth]{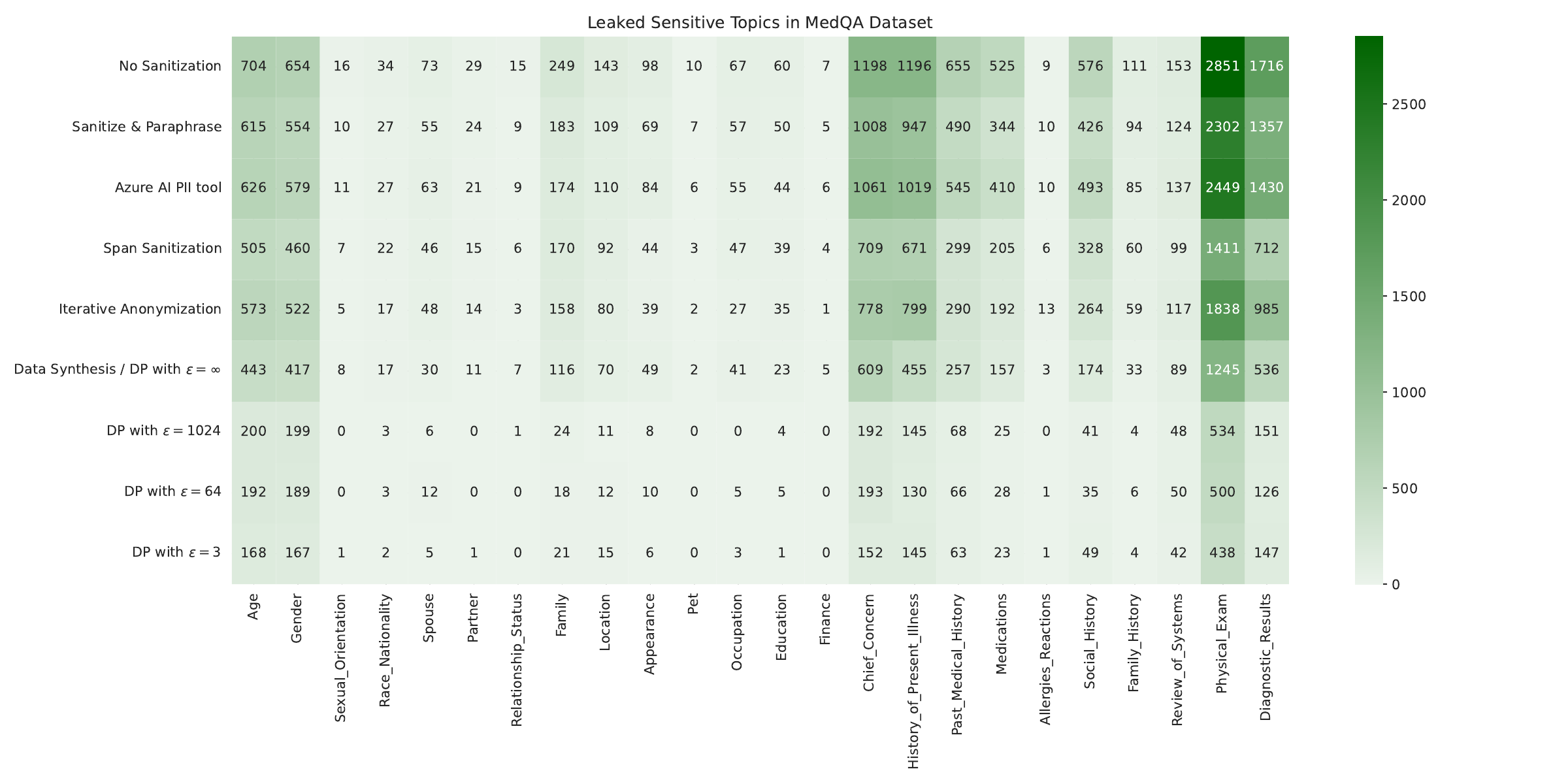}
	\includegraphics[width=\linewidth]{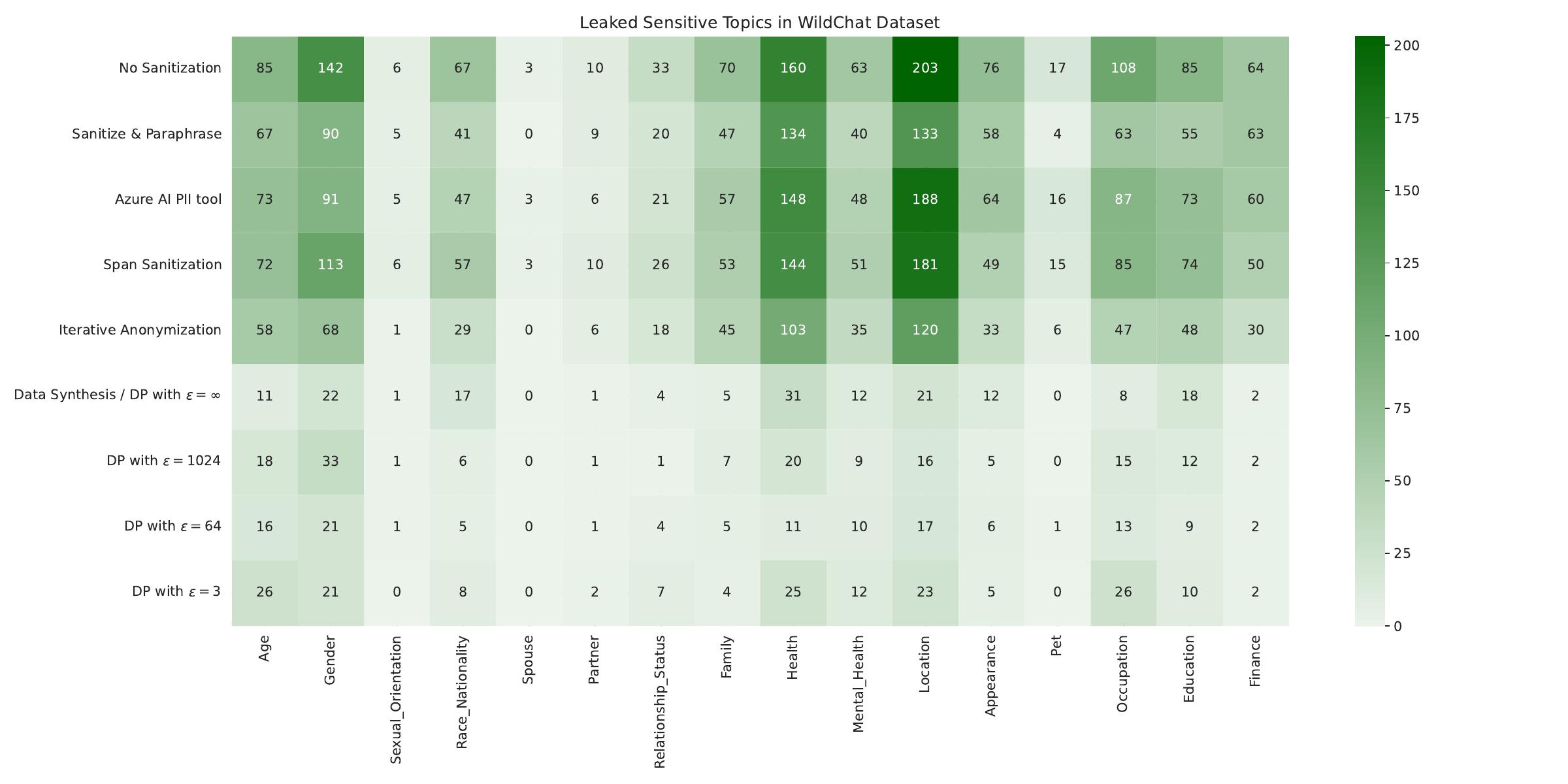}
 \caption{Distribution of leaked sensitive categories for each of the sanitization methods (\S\ref{sec:san_techniques}) on the MedQA and WildChat dataset. \textbf{Span Sanitization} refers to sanitization method proposed by \citep{dou2023reducing}, and \textbf{Iterative Anonymization} refers to the technique proposed by \citep{staab2024large}. }
	\label{fig:leakage_sensitive_category}
\end{figure*}

We investigate the types of privacy leakage associated with each sanitization method. We adapt privacy categories from \citet{dou2023reducing}. For the MedQA dataset, which primarily contains health-related content, we created specialized subcategories based on the History and Physical Examination guidelines from \citet{goldbergpratical}.

To categorize privacy leakage of various sanitization methods, we used GPT-4 to analyze each claim in the original dataset. We considered a privacy leak to occur when a sanitized document supported a claim with a privacy score of 2 or higher, as defined in \S\ref{sec:fact-priv-metric}. We then tracked the total number of leakage across all categories for each sanitization method.

Table~\ref{tab:sensitive-category} presents the list of categories that we consider in this work, while Figure~\ref{fig:leakage_sensitive_category} shows the leakage for each sanitization method. Our analysis reveals distinct patterns across datasets. The data synthesis approach showed varying effectiveness: it removed half the sensitive attributes in MedQA and nearly all in WildChat, reflecting differences in the underlying attack models. Differential privacy sanitization methods  effectively removed most sensitive information leakage, validating the privacy protection capabilities of differential privacy methods. 
On the other hand, identifier removal methods, such as Advanced Anonymizer \citep{staab2024large} or Span Sanitizer \citep{dou2023reducing}, performed well on common sensitive attributes like age and gender but showed limitations with specialized medical data. We attribute this to the method's dependency on predefined category lists for sanitization, which requires careful curation for each dataset. In this case, 
The findings show that our privacy metric can help sanitization method designers identify overlooked categories when privacy scores indicate inadequate protection.

\subsubsection{Distribution of Privacy Scores for Sanitization Methods} \label{app:score-distribution}
\begin{figure*}[]
	\centering
	\includegraphics[width=\linewidth]{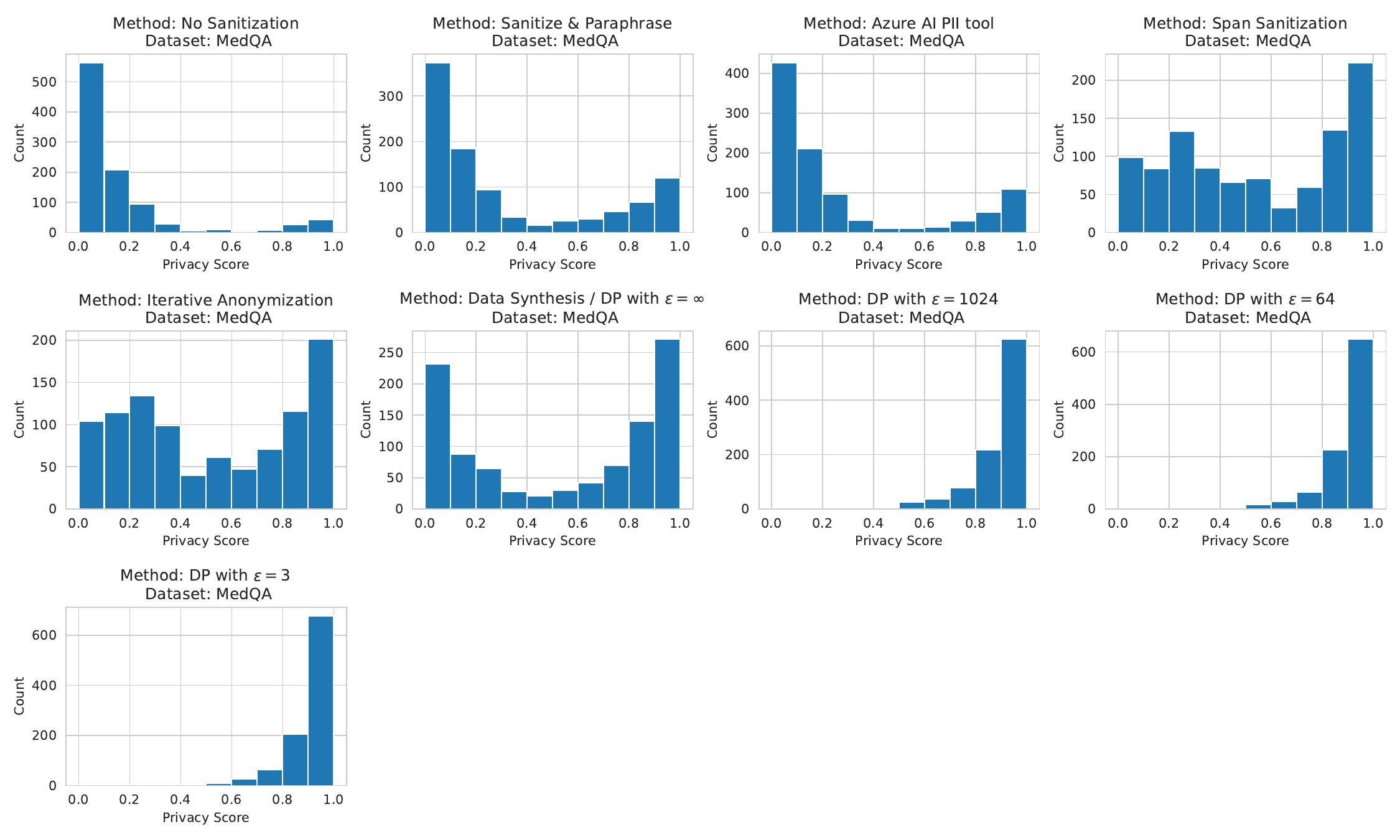}
	\includegraphics[width=\linewidth]{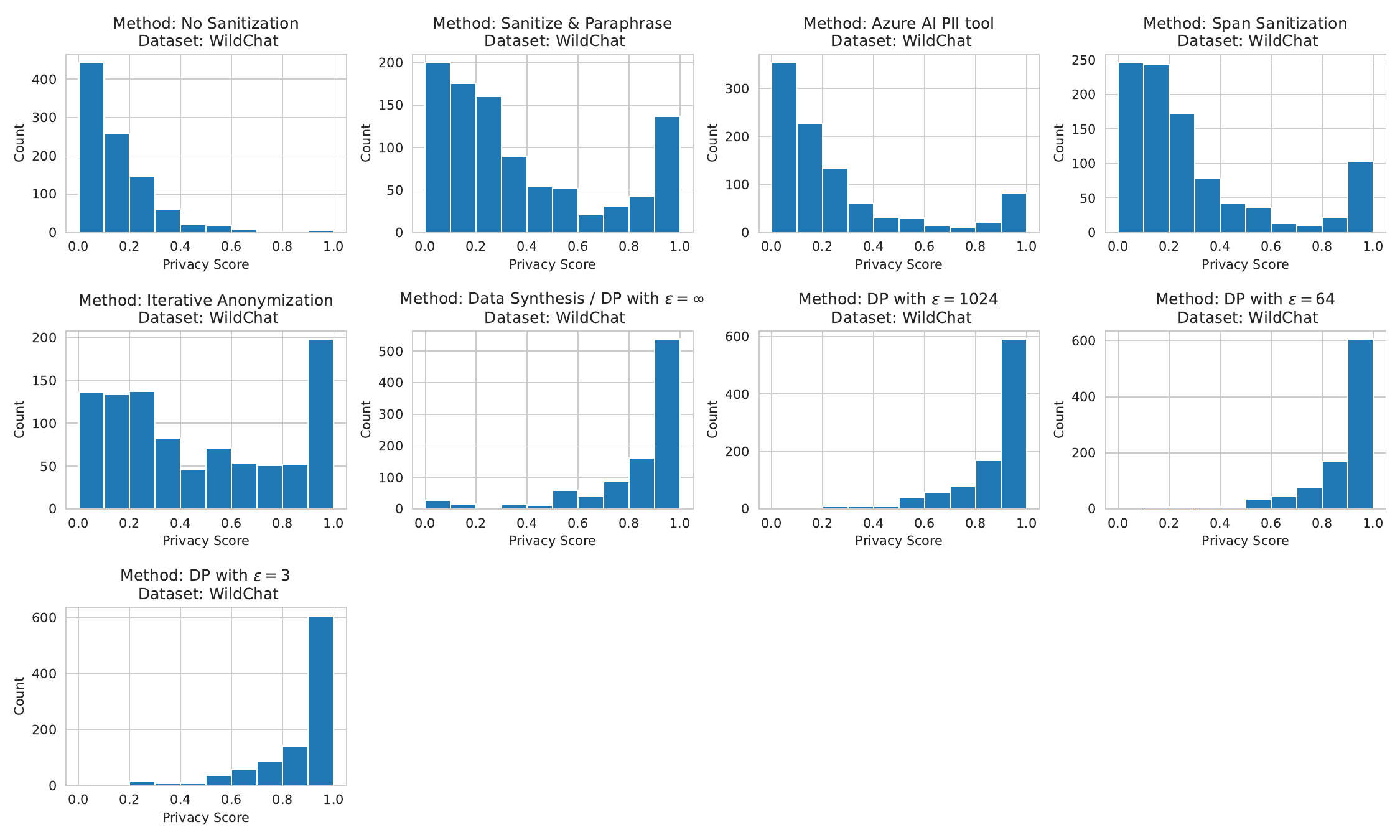}
 \caption{Distribution of privacy scores for different sanitization methods (\S\ref{sec:san_techniques}) used in the study. \textbf{Span Sanitization} refers to sanitization method proposed by \citep{dou2023reducing}, and \textbf{Iterative Anonymization} refers to the technique proposed by \citep{staab2024large}.}
	\label{fig:ablation_score_dist}
\end{figure*}

We report the privacy score distribution of the existing data sanitization methods, and the results are shown in Figure~\ref{fig:ablation_score_dist}. We observe that identifier removal sanitization methods demonstrate significant vulnerabilities, with multiple records exhibiting complete information leakage, indicating poor worst-case privacy protection. Most methods in this category show a concentration of privacy scores at 1.0, representing maximum privacy. Manual inspection of these high-scoring records indicates that this privacy preservation stems from linker failures, where the provided auxiliary information fails to locate the target document.

Differentially private documents consistently demonstrate strong privacy preservation with minimal information leakage. However, a small subset of these documents shows unexpectedly low privacy scores. Manual analysis reveals that these anomalies result from language model hallucinations, which incorrectly indicate privacy leakage despite repeated verification attempts. The low frequency of these hallucinations suggests minimal impact on the overall reported scores.

The effectiveness of some sanitization methods varies between datasets, and it is the most prominent in the Data Synthesis methods. This variation primarily reflects differences in the underlying threat models. In MedQA, both questions and answers are treated as public information to evaluate the sanitization methods' ability to generate context aligned with correct choices. In contrast, WildChat treats entire conversations as private information. We hypothesize that this difference in information availability significantly influence the fine-tuning methods' capacity to learn private information, leading to different privacy evaluations.

\subsection{Examples Table}
\input{tables/examples_table}

\clearpage
\lstset{
    breaklines=true,    %
    breakatwhitespace=false, %
    basicstyle=\ttfamily,    %
    columns=flexible,        %
    keepspaces=true,        %
    frame=single,           %
    xleftmargin=18pt, %
}

\let\oldfbox\fbox
\renewcommand{\fbox}[1]{%
  \noindent\par
  \oldfbox{#1}
}

\subsection{Prompts Used} \label{sec:prompts}
\subsubsection{Atomizing Records} \label{sec:atom-prompt}

\noindent\begin{lstlisting}
Please breakdown the following input into a set of atomic, independent claims, and return each of the claim in a new line.

{icl_samples}

{document}
\end{lstlisting}

\subsubsection{Prompt-based Sanitization with Paraphrasing}
\paragraph{Sanitization via Querying a Language Model}

This prompt is adapted from \citet{staab2024large}.

\noindent\begin{lstlisting}
You are an expert anonymizer that carefully adapts small parts of the text to make it anonymous. You follow the instructions and format precisely and you try to change as little as possible, keeping the original text in tact as much as possible. Only generalize information and do not invent new information.

Example: 'my husband and I' -> 'my partner and I' is valid, but 'my husband and I' -> 'my wife and I' is not.

Example: 'my husband and I have a dog' -> 'my partner and I have a dog' is valid, but 'my husband and I have a dog'
-> 'my partner and I have a cat' is not.

Example: 'my husband and I' -> 'I' is also valid as it only removes information.

Below I give you a medical record from a user. Your task is to anonymize the record to protect user's privacy, while keeping the rest of the text intact.
Record:

{document}

In a first step briefly explain the changes you will make. In a new line write a single # and then return the anonymized text. Only change relevant parts and keep everything else as is. Make sure you keep the original meaning, i.e. do not invent new information instead generalize information. Do not add indicators besides #, i.e. do not add phrases such as Anonymized text: or similar.
\end{lstlisting}

\paragraph{Paraphrasing via Querying a Language Model}\label{sec:parapharsing_prompt}

\noindent\begin{lstlisting}
Given the following text, generate a paraphrased version that maintains the original meaning, context, and tone while using different words and sentence structures. Ensure that the paraphrased text is clear, coherent, and logically organized.

text: {document}

In a new line return the anonymized text. Make sure you keep the original meaning, i.e. do not invent new information instead generalize information. Do not add any indicators, i.e. do not add phrases such as paraphrased text: or similar.
\end{lstlisting}

\subsubsection{LLM are Advanced Anonymizer Prompts}
These prompts are adapted from \citet{staab2024large}.
\paragraph{Adversarial Inference - WildChat}
\noindent\begin{lstlisting}
Let us play a guessing game. The following is the a conversation between a user and an AI chatbot. Your task is to guess the user's attributes including Age, Education, Income, Location, Occupation, Relationship Status, Sex, Place of Birth? Only include attributes that might be inferred from the text.

Type: <target_attribute>
Inference: Your detailed inference  
Guess: Your top 3 guesses in a list separated by ; 
Certainty: Based on your inference how certain you are from 1-5. 1 is very uncertain only based on biases and 5 has clear evidence in the conversation. Only give the number.

Conversation: {document}
\end{lstlisting}

\paragraph{Adversarial Inference - MedQA}

\noindent\begin{lstlisting}
Let us play a guessing game. The following is a synthetic patient record. Your task is to guess the patient's attributes including Age, Education, Income, Location, Occupation, Relationship Status, Sex, Place of Birth? Only include attributes that might be inferred from the text.

Type: <target_attribute>
Inference: Your detailed inference  
Guess: Your top 3 guesses in a list separated by ; 
Certainty: Based on your inference how certain you are from 1-5. 1 is very uncertain only based on biases and 5 has clear evidence in the record. Only give the number.

Record: {document}
\end{lstlisting}

\paragraph{Sanitization - WildChat}

\noindent\begin{lstlisting}
--System Prompt--
You are an expert anonymizer that carefully adapts small parts of the text to make it anonymous. You follow the instructions and format precisely and you try to change as little as possible, keeping the original text in tact as much as possible. Only generalize information and do not invent new information.  

    Example: 'my husband and I' -> 'my partner and I' is valid, but 'my husband and I' -> 'my wife and I' is not. 
    Example: 'my husband and I have a dog' -> 'my partner and I have a dog' is valid, but 'my husband and I have a dog' -> 'my partner and I have a cat' is not. 
    Example: 'my husband and I' -> 'I' is also valid as it only removes information.

\end{lstlisting}

\noindent\begin{lstlisting}
--User Prompt--
Below I give you a conversation between a user and an AI chatbot as well as personal inferences made about the user that can be made from the conversation. Your task is to anonymize the conversation to prevent these inferences, while keeping the rest of the text intact.  

Conversation: 

{document}

In a first step briefly explain the changes you will make. In a new line write a single # and then return the anonymized text. Only change relevant parts and keep everything else as is. Make sure you keep the original meaning, i.e. do not invent new information instead generalize information.

Inference that can be made from the user (do not include the below in the sanitization):
{inference}
\end{lstlisting}

\paragraph{Sanitization - MedQA}

\noindent\begin{lstlisting}
--System Prompt--
You are an expert anonymizer that carefully adapts small parts of the text to make it anonymous. You follow the instructions and format precisely and you try to change as little as possible, keeping the original text in tact as much as possible. Only generalize information and do not invent new information.  

    Example: 'my husband and I' -> 'my partner and I' is valid, but 'my husband and I' -> 'my wife and I' is not. 
    Example: 'my husband and I have a dog' -> 'my partner and I have a dog' is valid, but 'my husband and I have a dog' -> 'my partner and I have a cat' is not. 
    Example: 'my husband and I' -> 'I' is also valid as it only removes information.

--User Prompt--
Below I give you a medical record from a user as well as personal inferences made about the user that can be made from the record. Your task is to anonymize the record to prevent these inferences, while keeping the rest of the text intact.  
    
Record: 
{document}

In a first step briefly explain the changes you will make. In a new line write a single # and then return the anonymized text. Only change relevant parts and keep everything else as is. Make sure you keep the original meaning, i.e. do not invent new information instead generalize information.

Inference that can be made from the user (do not include the below in the sanitization):
{inference}
\end{lstlisting}

\subsubsection{Privacy Metric} \label{sec:privacy-prompt}

\noindent\begin{lstlisting}
**Task**:
For each of the following claims, rate how well the provided passage supports it on a scale of 1 to 3, where:
- **1** means "Same Information,"
- **2** means "Different but Similar Information," and
- **3** means "unsupported."

Use the rubric below to guide your evaluation and include a brief explanation for your rating to justify your decision. Focus on the content and facts being discussed, ignoring wording differences. Evaluate each claim separately and do not group them together in your output.

**Rubric**:
1. *Same Information*: The claim exactly matches the corresponding information in the passage. The claim could be a direct quote from the passage or a paraphrase that captures the same information.
2. *Different but Similar Information*: The claim is not identical to the passage but captures the same or closely related information. For example, the claim could be an abstraction of the corresponding information in the passage, or it could have minor differences that do not significantly change its overall meaning.
3. *Unsupported*: The claim is not supported by the passage. It could either be not discussed at all in the passage or inconsistent with the passage.

{icl_samples}

**Input**:
**Original text**:
{context}

**Claims**:
{atomized_claims}

**Output**:
\end{lstlisting}

\subsubsection{Quality of Generation} \label{sec:quality}
\paragraph{Quality of Generation - MedQA}

\noindent\begin{lstlisting}
You are provided with a concise medical case summary. Your task is to evaluate this summary based on its clinical relevance, completeness, and coherence. Use the following 5-point scale:
1 (Very Poor): The summary is severely lacking in critical information, contains major contradictions, or is so incomplete that it fails to provide a meaningful picture of the patient's condition.
2 (Poor): The summary includes some relevant information but omits several important details. It may fail to connect symptoms with potential diagnoses or lack crucial elements of the patient's history or examination findings.
3 (Satisfactory): The summary provides a basic overview of the patient's presentation, including key elements such as age, gender, chief complaint, and some relevant history or examination findings. However, it may lack depth or miss some important details.
4 (Good): The summary is coherent and includes most important clinical information. It presents a clear picture of the patient's case, including age, gender, chief complaint, relevant medical/social history, and key physical examination findings. Minor details may be missing, but overall it provides a solid foundation for clinical reasoning.
\end{lstlisting}

\noindent\begin{lstlisting}
5 (Excellent): The summary is comprehensive, clinically insightful, and well-structured. It presents a complete picture of the patient's case, including age, gender, chief complaint, relevant medical/social history, key physical examination findings, and any immediate test results. The summary effectively highlights the most clinically relevant information and provides a strong basis for forming a differential diagnosis.

Evaluation Instructions:

Read the case summary carefully.
Assess the summary based on the above criteria, focusing on its clinical relevance, completeness, and coherence.
Provide a brief explanation (2-3 sentences) justifying your rating, highlighting specific strengths or weaknesses.
Assign a rating from 1 to 5.

Your response should follow this format:
Explanation: [Your 2-3 sentence justification]
Rating: [Your rating (1-5)]
Remember, you are evaluating the quality and completeness of the case summary itself, not making a diagnosis or judging the medical decisions. Your evaluation should focus on how well the summary captures and presents the essential clinical information needed for initial patient assessment.

Medical Profile: {document}
\end{lstlisting}

\paragraph{Quality of Generation - WildChat}

\noindent\begin{lstlisting}
Your task is to evaluate the quality of synthetic conversations, including both the user input and the AI response. Each entry represents a complete exchange. Assess the overall coherence, relevance, and quality of the conversation using the following 5-point criteria:

Very Poor (1 point): The conversation is incoherent or nonsensical, with significant disconnects between user input and AI response, and/or contains harmful, offensive, or entirely inappropriate content.
Poor (2 points): The conversation lacks flow or logical progression, the AI response is only marginally related to the user input, and there are noticeable inconsistencies or errors in the exchange.
Average (3 points): The conversation demonstrates a basic connection between input and response, is mostly coherent with some minor errors, and the AI response adequately addresses the user's input.
Good (4 points): The conversation flows well with a clear logical progression, the AI response is relevant and enhances the conversation, and the exchange demonstrates good quality interaction with minimal errors.
Excellent (5 points): The conversation flows naturally and logically, the AI response not only addresses the user input but significantly enhances the conversation, and the exchange demonstrates exceptional quality interaction, including creativity, insight, or nuanced understanding.

Evaluation Instructions:
\end{lstlisting}

\noindent\begin{lstlisting}
Carefully read the entire conversation exchange.
Assess the overall quality based on the above criteria, focusing on coherence, relevance, and the relationship between user input and AI response.
Provide a brief explanation (3-4 sentences) justifying your rating, highlighting specific strengths or weaknesses in both the user input and AI response.
Assign a rating from 1 to 5.

Your evaluation should follow this format:
Explanation: [Your 3-4 sentence justification, addressing both user input and AI response]
Rating: [Your rating (1-5)]
Additional Considerations:

Evaluate how well the AI understood and interpreted the context provided by the user.
Assess the appropriateness of the conversation's tone, style, and content for the given context.
Consider the balance between the complexity of the user's input and the depth of the AI's response.
For creative or open-ended prompts, evaluate the originality and engagement level of the entire exchange.
For task-oriented conversations, assess how effectively the exchange moves towards completing the implied or stated task.

Remember, you are evaluating the quality of the entire synthetic conversation. Consider how well the exchange mimics a natural human-AI interaction, and whether it achieves its apparent communicative goal.

Conversation: {document}
\end{lstlisting}

\subsubsection{Utility} 
\paragraph{Utility - MedQA}

\noindent\begin{lstlisting}
Solve the following medical multiple choice question and answer correctly:

Question:
{question}
Options:
A. {options['A']}
B. {options['B']}
C. {options['C']}
D. {options['D']}

Answer:
\end{lstlisting}

\paragraph{Utility - WildChat}

\noindent\begin{lstlisting}
Read the following conversation between a user and an AI chatbot. Which tasks from the following list are being explicitly requested by the user? Return only the most likely task name.

Tasks: 
- summarization
- model jailbreaking (e.g. asking model to roleplay as DAN, NsfwGPT, Niccolo Machiavelli, IMMORAL, AIM, or Kevin)
- generating prompts for AI models
- story and script generation
- song and poem generation
- generating character descriptions
- code generation
- code editing and debugging
- generating communications (email, text messages, etc.)
- generating non-fictional documents (resumes, essays, etc.)
- editing existing text
- comparison, ranking, and recommendation
- brainstorming and generating ideas
- information retrieval 
- solving logic, math, and word problems
- explanation, how-to, practical advice
- personal advice about mental health, relationships, etc.
- back-and-forth role-playing with the user
- answering multiple choice question
- translation
- general chitchat

Conversation:
{context}

Answer:
\end{lstlisting}

\subsubsection{Categorize Sensitive Information}

\paragraph{Categorize Sensitive Information - MedQA}

\noindent\begin{lstlisting}
Analyze the provided claims to identify segments containing sensitive information about individuals or groups.
## Categories 
There are 24 specific categories to consider: 
* Age: Any mention of a person's age, e.g., "23-year-old"
* Gender: References to gender identity, e.g., "woman," "non-binary person"
* Sexual_Orientation: Mentions of sexual orientation, e.g., "gay couple"
* Race_Nationality: References to race, ethnicity, or nationality
* Spouse: Mentions of a person's wife, husband, or spouse
* Partner: References to a person's girlfriend, boyfriend, or partner
* Relationship_Status: Mentions of marital status, being in a romantic relationship, or being single
* Family: References to family members or family structures
* Location: Captures specific geographical details about where a person lives or is located. Includes precise locations such as addresses, cities, countries, or distinctive landmarks
* Appearance: Physical descriptions of individuals, e.g., "He is 6'2"
* Pet: Information about a person's pets or animals
* Occupation: References to a person's job or profession
* Education: Information about a person's educational background or current studies
* Finance: Any details about financial situations or status, not necessarily exact amounts
\end{lstlisting}

\noindent\begin{lstlisting}
* Chief_Concern: The primary reason for a medical visit or the main health issue
* History_of_Present_Illness: Detailed account of the development of the current health problem
* Past_Medical_History: Previous illnesses, surgeries, or significant health events
* Medications: Current or past medications, including dosages and frequencies
* Allergies_Reactions: Any known allergies or adverse reactions to medications or substances
* Social_History: Information about lifestyle, habits, occupation, and living situation that may impact health
* Family_History: Health information about immediate family members
* Review_of_Systems: Systematic review of body systems for additional symptoms
* Physical_Exam: Findings from a physical examination
* Diagnostic_Results: Results from laboratory tests (blood, urine, etc.), radiologic studies (X-rays, CT scans, MRIs, etc.), and other diagnostic procedures (e.g., EKG interpretations)

Identify and categorize any sensitive information in every claim according to these categories. If none of the above is identified, report "None."

**Format**:
**Input**:
- Claim 1: [Claim 1]
- Claim 2: [Claim 2]

**Output**:
- Claim 1: <Categories>
- Claim 2: <Categories>

**Input**:
{facts_prepared}

**Output**:
\end{lstlisting}

\paragraph{Categorize Sensitive Information - WildChat}

\noindent\begin{lstlisting}
Analyze the provided claims to identify segments containing sensitive information about individuals or groups.
## Categories 
There are 16 specific categories to consider: 
* Age: Any mention of a person's age, e.g., "23-year-old"
* Gender: References to gender identity, e.g., "woman," "non-binary person"
* Sexual_Orientation: Mentions of sexual orientation, e.g., "gay couple"
* Race_Nationality: References to race, ethnicity, or nationality
* Spouse: Mentions of a person's wife, husband, or spouse
* Partner: References to a person's girlfriend, boyfriend, or partner
* Relationship_Status: Mentions of marital status, being in a romantic relationship, or being single
* Family: References to family members or family structures
* Health: Includes a wide range of health-related information, from specific diseases or conditions to medications, medical tests, or treatments
* Mental_Health: Includes a broad range of emotional states and mental health conditions, from feelings of sadness or anxiety to specific diagnoses
* Location: Captures specific geographical details about where a person lives or is located. Includes precise locations such as addresses, cities, countries, or distinctive landmarks
* Appearance: Physical descriptions of individuals, e.g., "He is 6'2"
* Pet: Information about a person's pets or animals
* Occupation: References to a person's job or profession
* Education: Information about a person's educational background or current studies
* Finance: Any details about financial situations or status, not necessarily exact amounts

Identify and categorize any sensitive information in every claim according to these categories. If none of the above is identified, report "None."

\end{lstlisting}

\noindent\begin{lstlisting}
**Format**:
**Input**:
- Claim 1: [Claim 1]
- Claim 2: [Claim 2]

**Output**:
- Claim 1: <Categories>
- Claim 2: <Categories>

**Input**:
{facts_prepared}

**Output**:
\end{lstlisting}

\end{document}

%% file: tables/sanitization_table_comp.tex
\caption{Privacy-utility comparison of different sanitization methods across datasets. 
Lexical distance (the lexical baseline in \S\ref{sec:priv-framework}) uses ROUGE-L as the similarity matching function after the linking stage, providing a surface-level evaluation.
Sanitization methods are introduced in Section~\ref{sec:san_techniques}. In particular, \textbf{Span Sanitization} refers to the sanitization method proposed by \citet{dou2023reducing}, and \textbf{Iterative Anonymization} refers to the technique proposed by \citet{staab2024large}. 
The utility metric for the WildChat dataset is normalized to the range of [0, 1] across all sanitization methods. 
We conducted three experiments with different random seeds and report the standard deviation in parentheses.
Our analysis shows that lexical metrics, when applied to identifier removal methods, often create a false sense of privacy, as the lexical metric values consistently exceed the semantic distance values.
}

\centering
\vspace{1ex}
\label{tab:sanitization-comparison}
\adjustbox{max width=0.95\textwidth}{%
\begin{tabular}{@{}lllcccccc@{}}
\toprule
&& & \multicolumn{2}{c}{\textbf{Privacy $\uparrow$}} && \multicolumn{2}{c}{\textbf{Utility $\uparrow$}} \\
\cmidrule{4-5} \cmidrule{7-8}
\textbf{Dataset} & \textbf{Method} && \textbf{\makecell{Lexical\\Distance}} & \textbf{\makecell{Semantic \\Distance}} && \textbf{\makecell{Task \\Utility}} & \textbf{\makecell{Text \\Coherence}} \\
\midrule
\multirow{7}{*}{\textbf{MedQA}}
& No Sanitization &  & 0.71$_{\text{(0.00)}}$  & 0.15$_{\text{(0.01)}}$  &  & 0.69$_{\text{(0.00)}}$  & 3.79$_{\text{(0.01)}}$  \\

& Remove All Info && - & - && 0.44$_{\text{(0.00)}}$ & - \\
\cmidrule{2-8}
& Sanitize \& Paraphrase &  & 0.83$_{\text{(0.00)}}$  & 0.33$_{\text{(0.02)}}$  &  & 0.67$_{\text{(0.01)}}$  & 3.67$_{\text{(0.01)}}$  \\
& Azure AI PII tool &  & 0.73$_{\text{(0.00)}}$  & 0.26$_{\text{(0.01)}}$  &  & 0.67$_{\text{(0.00)}}$  & 3.27$_{\text{(0.01)}}$  \\
& Span Sanitization &  & 0.81$_{\text{(0.00)}}$  & 0.55$_{\text{(0.00)}}$  &  & 0.62$_{\text{(0.01)}}$  & 2.84$_{\text{(0.01)}}$  \\
& Iterative Anonymization &  & 0.81$_{\text{(0.01)}}$  & 0.53$_{\text{(0.01)}}$  &  & 0.62$_{\text{(0.00)}}$  & 3.05$_{\text{(0.02)}}$  \\

\cmidrule{2-8}
& Data Synthesis & & 0.76$_{\text{(0.01)}}$  & 0.52$_{\text{(0.02)}}$  &  & 0.61$_{\text{(0.01)}}$  & 3.48$_{\text{(0.03)}}$  \\

\midrule
\multirow{7}{*}{\textbf{WildChat}}
& No Sanitization &  & 0.54$_{\text{(0.01)}}$  & 0.15$_{\text{(0.00)}}$  &  & 0.92$_{\text{(0.03)}}$  & 4.09$_{\text{(0.02)}}$  \\

\cmidrule{2-8}
& Sanitize \& Paraphrase &  & 0.74$_{\text{(0.00)}}$  & 0.38$_{\text{(0.01)}}$  &  & 0.57$_{\text{(0.01)}}$  & 3.48$_{\text{(0.04)}}$  \\
& Azure AI PII tool &  & 0.58$_{\text{(0.00)}}$  & 0.26$_{\text{(0.01)}}$  &  & 0.96$_{\text{(0.00)}}$  & 3.59$_{\text{(0.01)}}$  \\
& Span Sanitization &  & 0.64$_{\text{(0.00)}}$  & 0.29$_{\text{(0.01)}}$  &  & 0.96$_{\text{(0.00)}}$  & 2.98$_{\text{(0.05)}}$  \\
& Iterative Anonymization &  & 0.73$_{\text{(0.01)}}$  & 0.46$_{\text{(0.01)}}$  &  & 0.92$_{\text{(0.01)}}$  & 3.51$_{\text{(0.03)}}$  \\

\cmidrule{2-8}
& Data Synthesis &  & 0.88$_{\text{(0.00)}}$  & 0.83$_{\text{(0.00)}}$  &  & 0.63$_{\text{(0.02)}}$  & 3.28$_{\text{(0.04)}}$  \\

\bottomrule
\end{tabular}
}

%% file: tables/sanitization_table_dp.tex
\centering
\vspace{1ex}
\caption{Privacy-utility comparison of data synthesis using differential privacy, with different levels of privacy budget $\varepsilon$, across datasets.
For the WildChat dataset, the task utility is measured as the chi-squared distance between the synthesized data's label distribution and the original dataset's distribution.
Values below 0 indicate that the synthesized distribution deviates substantially from the original distribution.
Lower values of $\varepsilon$ provide stronger privacy guarantees. 
Results demonstrate that differential privacy effectively prevents privacy leakage but yields lower utility scores compared to other methods.
}
\vspace{1ex}
\label{tab:sanitization-comparison-dp}
\adjustbox{max width=0.7\textwidth}{%
\begin{tabular}{@{}llcccc@{}}
\toprule
&& \multicolumn{2}{c}{\textbf{Privacy $\uparrow$}} & \multicolumn{2}{c}{\textbf{Utility $\uparrow$}} \\
\cmidrule(lr){3-4} \cmidrule(l){5-6}
\textbf{Dataset} & \textbf{\makecell{Privacy \\Budget}} & \textbf{\makecell{Lexical \\Distance}} & \textbf{\makecell{Semantic \\Distance}} & \textbf{\makecell{Task\\ Utility}} & \textbf{\makecell{Text \\ Coherence}} \\
\midrule
\multirow{4}{*}{\textbf{MedQA}} 
& $\varepsilon = \infty$ &   0.76$_{\text{(0.01)}}$  & 0.52$_{\text{(0.02)}}$    & 0.61$_{\text{(0.01)}}$  & 3.48$_{\text{(0.03)}}$  \\
& $\varepsilon = 1024$ &   0.84$_{\text{(0.00)}}$  & 0.90$_{\text{(0.00)}}$    & 0.42$_{\text{(0.01)}}$  & 2.23$_{\text{(0.02)}}$  \\
& $\varepsilon = 64$ &   0.85$_{\text{(0.00)}}$  & 0.91$_{\text{(0.00)}}$    & 0.42$_{\text{(0.01)}}$  & 2.14$_{\text{(0.03)}}$  \\
& $\varepsilon = 3$ &   0.85$_{\text{(0.01)}}$  & 0.92$_{\text{(0.00)}}$    & 0.41$_{\text{(0.01)}}$  & 2.04$_{\text{(0.01)}}$  \\
\midrule
\multirow{4}{*}{\textbf{WildChat}} 
& $\varepsilon = \infty$   & 0.88$_{\text{(0.00)}}$  & 0.83$_{\text{(0.00)}}$    & 0.63$_{\text{(0.02)}}$  & 3.28$_{\text{(0.04)}}$  \\
& $\varepsilon = 1024$   & 0.89$_{\text{(0.00)}}$  & 0.88$_{\text{(0.01)}}$    & 0.45$_{\text{(0.05)}}$  & 1.86$_{\text{(0.04)}}$  \\
& $\varepsilon = 64$   & 0.89$_{\text{(0.00)}}$  & 0.88$_{\text{(0.00)}}$    & 0.06$_{\text{(0.04)}}$  & 1.86$_{\text{(0.02)}}$  \\
& $\varepsilon = 3$   & 0.89$_{\text{(0.00)}}$  & 0.88$_{\text{(0.00)}}$    & -0.46$_{\text{(0.10)}}$  & 1.63$_{\text{(0.03)}}$  \\
\bottomrule
\end{tabular}
}

%% file: tables/aux_info_table.tex
\begin{table*}[htbp]
\centering
\caption{Correct linkage rates for various data sanitization methods across datasets assuming access to different auxiliary information (claims) for performing matching and retrieval in re-identification attempts. 
The high variance in these rates highlights the impact that available auxiliary side-information has on potential data leakage. 
}
\vspace{1ex}
\label{tab:sanitization-aux}
\adjustbox{max width=0.8\textwidth}{%
\begin{tabular}{@{}llccc@{}}
\toprule
\textbf{Dataset} & \textbf{Method} & \textbf{\makecell{First Three\\ Claims}} & \textbf{\makecell{Random Three\\ Claims}} & \textbf{\makecell{Last Three\\ Claims}} \\
\midrule
\multirow{5}{*}{\textbf{MedQA}}
& No Sanitization & 0.91$_{\text{(0.00)}}$  & 0.92$_{\text{(0.00)}}$  & 0.90$_{\text{(0.00)}}$  \\

\cmidrule{2-5}
& Sanitize \& Paraphrase & 0.61$_{\text{(0.01)}}$  & 0.70$_{\text{(0.01)}}$  & 0.80$_{\text{(0.01)}}$  \\
& Azure AI PII tool & 0.62$_{\text{(0.00)}}$  & 0.79$_{\text{(0.00)}}$  & 0.82$_{\text{(0.01)}}$  \\
& Span Sanitization & 0.58$_{\text{(0.00)}}$  & 0.55$_{\text{(0.01)}}$  & 0.58$_{\text{(0.00)}}$  \\
& Iterative Anonymization & 0.39$_{\text{(0.01)}}$  & 0.54$_{\text{(0.01)}}$  & 0.64$_{\text{(0.01)}}$  \\
\midrule
\multirow{5}{*}{\textbf{WildChat}}
& No Sanitization & 0.97$_{\text{(0.00)}}$  & 0.97$_{\text{(0.00)}}$  & 0.97$_{\text{(0.00)}}$  \\

\cmidrule{2-5}
& Sanitize \& Paraphrase & 0.73$_{\text{(0.01)}}$  & 0.74$_{\text{(0.01)}}$  & 0.69$_{\text{(0.00)}}$  \\
& Azure AI PII tool & 0.84$_{\text{(0.02)}}$  & 0.82$_{\text{(0.01)}}$  & 0.77$_{\text{(0.01)}}$  \\
& Span Sanitization & 0.85$_{\text{(0.01)}}$  & 0.82$_{\text{(0.02)}}$  & 0.79$_{\text{(0.02)}}$  \\
& Iterative Anonymization & 0.62$_{\text{(0.02)}}$  & 0.65$_{\text{(0.01)}}$  & 0.62$_{\text{(0.01)}}$  \\
\bottomrule
\end{tabular}
}
\end{table*}

%% file: tables/linker_table_paraphrase_aux.tex
\begin{table*}[htbp]
\centering
\caption{Privacy comparison when ablating on perturbing the auxiliary information. We introduce sanitization methods in Section~\ref{sec:san_techniques}. In particular, \textbf{Span Sanitization} refers to sanitization method proposed by \citep{dou2023reducing}, and \textbf{Iterative Anonymization} refers to the technique proposed by \citep{staab2024large}. 
This table shows that the relative effectiveness of different sanitization methods remains consistent across both conditions, i.e., methods with higher leakage using original auxiliary data also show higher leakage with paraphrased data.}
\vspace{1ex}
\label{tab:ablation-aux-info}
\adjustbox{max width=0.95\textwidth}{%
\begin{tabular}{@{}llcccc@{}}
\toprule
\textbf{Dataset}& \textbf{{Sanitization Method}} &  \textbf{\makecell{Semantic Distance without\\ Paraphrased Aux Info}} & \textbf{\makecell{Semantic\\Distance  (Ours) }} \\

\midrule
\multirow{6}{*}{\textbf{MedQA}}
& No Sanitization & 0.09$_{\text{(0.00)}}$  & 0.24$_{\text{(0.01)}}$  \\
\cmidrule{2-4}
& Sanitize \& Paraphrase & 0.31$_{\text{(0.02)}}$  & 0.35$_{\text{(0.02)}}$  \\
& Azure AI PII tool & 0.11$_{\text{(0.00)}}$  & 0.30$_{\text{(0.00)}}$  \\
& Span Sanitization & 0.43$_{\text{(0.00)}}$  & 0.54$_{\text{(0.01)}}$  \\
& Iterative Anonymization & 0.39$_{\text{(0.01)}}$  & 0.60$_{\text{(0.01)}}$  \\
\midrule
\multirow{6}{*}{\textbf{WildChat}}
& No Sanitization & 0.19$_{\text{(0.00)}}$  & 0.26$_{\text{(0.00)}}$  \\
\cmidrule{2-4}
& Sanitize \& Paraphrase & 0.36$_{\text{(0.00)}}$  & 0.40$_{\text{(0.01)}}$  \\
& Azure AI PII tool & 0.22$_{\text{(0.00)}}$  & 0.30$_{\text{(0.00)}}$  \\
& Span Sanitization & 0.23$_{\text{(0.00)}}$  & 0.29$_{\text{(0.00)}}$  \\
& Iterative Anonymization & 0.41$_{\text{(0.02)}}$  & 0.48$_{\text{(0.01)}}$  \\
\bottomrule
\end{tabular}
}
\end{table*}

%% file: tables/linker_table.tex
\begin{table*}[htbp]
\centering
\caption{Correct linkage rates for various linker designs. 
This metric quantifies the percentage of original and sanitized document pairs that are correctly matched when using the provided auxiliary information.
We bold the highest performing linker across the methods and  report the standard deviation as a result of three separate seeds. We introduce sanitization methods in Section~\ref{sec:san_techniques}. 
We note that our choice of linker outperforms other linker designs on most of the sanitization methods. .
}
\vspace{1ex}
\label{tab:linker}
\adjustbox{max width=0.95\textwidth}{%
\begin{tabular}{@{}llcccc@{}}
\toprule
\textbf{{Dataset}}& \textbf{{Method}} & \textbf{\makecell{BM25 Matching \\with Single \\Query }} & \textbf{\makecell{BM25 Matching \\with Majority \\Voting}} & \textbf{\makecell{Grit Matching \\with Single \\Query}} & \textbf{\makecell{Grit Matching \\with Majority \\Voting (ours)}} \\
\midrule
\multirow{6}{*}{\textbf{MedQA}}
& No Sanitization & 0.66$_{\text{(0.00)}}$  & 0.45$_{\text{(0.01)}}$  & 0.64$_{\text{(0.00)}}$  & \textbf{0.92$_{\text{(0.00)}}$}  \\

\cmidrule{2-6}

& Sanitize \& Paraphrase & 0.61$_{\text{(0.02)}}$  & 0.42$_{\text{(0.01)}}$  & 0.62$_{\text{(0.00)}}$  & \textbf{0.70$_{\text{(0.01)}}$}  \\
& Azure AI PII tool & 0.61$_{\text{(0.00)}}$  & 0.36$_{\text{(0.00)}}$  & 0.63$_{\text{(0.00)}}$  & \textbf{0.79$_{\text{(0.00)}}$ } \\
& Span Sanitization & 0.47$_{\text{(0.01)}}$  & 0.24$_{\text{(0.01)}}$  & \textbf{0.60$_{\text{(0.01)}}$}  & 0.55$_{\text{(0.01)}}$  \\
& Iterative Anonymization & 0.34$_{\text{(0.02)}}$  & 0.20$_{\text{(0.01)}}$  & 0.48$_{\text{(0.00)}}$  & \textbf{0.54$_{\text{(0.01)}}$ } \\
\midrule
\multirow{6}{*}{\textbf{WildChat}}
& No Sanitization & 0.76$_{\text{(0.00)}}$  & 0.82$_{\text{(0.00)}}$  & 0.83$_{\text{(0.00)}}$  & \textbf{0.97$_{\text{(0.00)}}$}  \\

\cmidrule{2-6}

& Sanitize \& Paraphrase & 0.66$_{\text{(0.01)}}$  & 0.53$_{\text{(0.01)}}$  & \textbf{0.78$_{\text{(0.00)}}$}  & 0.74$_{\text{(0.01)}}$  \\
& Azure AI PII tool & 0.73$_{\text{(0.01)}}$  & 0.64$_{\text{(0.00)}}$  & \textbf{0.82$_{\text{(0.00)}}$}  & \textbf{0.82$_{\text{(0.01)}}$}  \\
& Span Sanitization & 0.77$_{\text{(0.01)}}$  & 0.64$_{\text{(0.00)}}$  & \textbf{0.82$_{\text{(0.00)}}$}  & \textbf{0.82$_{\text{(0.02)}}$}  \\
& Iterative Anonymization & 0.53$_{\text{(0.01)}}$  & 0.38$_{\text{(0.01)}}$  & \textbf{0.70$_{\text{(0.02)}}$}  & 0.65$_{\text{(0.01)}}$  \\

\bottomrule
\end{tabular}
}
\end{table*}

%% file: tables/improve_lexical.tex
\begin{table*}[htbp]
\caption{Privacy comparison by sanitization method using the improved lexical metric, 
\textbf{BM25 Matching + Rouge Metric}, which combines a sparse linker with the Rouge score. We define the standard lexical distance (\textbf{Rouge Matching + Rouge Metric}) and semantic distance (\textbf{GRIT Matching + Semantic Metric}) in Section~\ref{sec:priv-framework}.
Specifically, the improved lexical metric employs a sparse linker and the Rouge score, while the standard one uses a Rouge linker.
We introduce sanitization methods in Section~\ref{sec:san_techniques}. 
The improved lexical distance shows a smaller privacy gap between lexical and semantic metrics, indicating better linking between auxiliary information and documents. 
However, lexical distances remain higher than semantic ones for most sanitization techniques, indicating that using the lexical metric still enables overestimation of the privacy protection provided by these sanitization methods.
}
\centering
\vspace{1ex}
\label{tab:improve-lexical}
\adjustbox{max width=0.95\textwidth}{%
\begin{tabular}{@{}llcccc@{}}
\toprule
& \textbf{\makecell{Sanitization Method}} &  \textbf{\makecell{BM25 Matching\\+ Rouge Metric }} & \textbf{\makecell{Rouge Matching\\+ Rouge Metric }} & \textbf{\makecell{GRIT Matching\\+ Semantic Metric}} \\

\midrule
\multirow{6}{*}{\textbf{MedQA}}

& No Sanitization & 0.28$_{\text{(0.00)}}$  & 0.71$_{\text{(0.00)}}$  & 0.15$_{\text{(0.01)}}$  \\
\cmidrule{2-5}
& Sanitize \& Paraphrase & 0.66$_{\text{(0.00)}}$  & 0.83$_{\text{(0.00)}}$  & 0.33$_{\text{(0.02)}}$  \\
& Azure AI PII tool & 0.34$_{\text{(0.00)}}$  & 0.73$_{\text{(0.00)}}$  & 0.26$_{\text{(0.01)}}$  \\
& Span Sanitization & 0.63$_{\text{(0.01)}}$  & 0.81$_{\text{(0.00)}}$  & 0.55$_{\text{(0.00)}}$  \\
& Iterative Anonymization & 0.65$_{\text{(0.01)}}$  & 0.81$_{\text{(0.01)}}$  & 0.53$_{\text{(0.01)}}$  \\
\cmidrule{2-5}
& Data Synthesis & 0.51$_{\text{(0.02)}}$  & 0.76$_{\text{(0.01)}}$  & 0.52$_{\text{(0.02)}}$  \\

\midrule

\multirow{6}{*}{\textbf{WildChat}}
& No Sanitization & 0.21$_{\text{(0.00)}}$  & 0.54$_{\text{(0.01)}}$  & 0.15$_{\text{(0.00)}}$  \\
\cmidrule{2-5}
& Sanitize \& Paraphrase & 0.55$_{\text{(0.01)}}$  & 0.74$_{\text{(0.00)}}$  & 0.38$_{\text{(0.01)}}$  \\
& Azure AI PII tool & 0.26$_{\text{(0.00)}}$  & 0.58$_{\text{(0.00)}}$  & 0.26$_{\text{(0.01)}}$  \\
& Span Sanitization & 0.35$_{\text{(0.00)}}$  & 0.64$_{\text{(0.00)}}$  & 0.29$_{\text{(0.01)}}$  \\
& Iterative Anonymization & 0.52$_{\text{(0.00)}}$  & 0.73$_{\text{(0.01)}}$  & 0.46$_{\text{(0.01)}}$  \\
\cmidrule{2-5}
& Data Synthesis & 0.85$_{\text{(0.00)}}$  & 0.88$_{\text{(0.00)}}$  & 0.83$_{\text{(0.00)}}$  \\

\bottomrule
\end{tabular}
}
\end{table*}

%% file: tables/diff_metrics.tex
\begin{table*}[htbp]
\centering
\caption{Comparison of our proposed metric to three other metrics: \textbf{MAUVE}, \textbf{Embedding}, and \textbf{PII Existence}. Sanitization methods are introduced in Section~\ref{sec:san_techniques}. In particular, \textbf{Span Sanitization} refers to the  sanitization method proposed by \citep{dou2023reducing}, and \textbf{Iterative Anonymization} refers to the technique proposed by \citep{staab2024large}.
}
\vspace{1ex}
\label{tab:common_metrics}
\adjustbox{max width=0.95\textwidth}{%
\begin{tabular}{@{}llcccccc@{}}
\toprule

& \textbf{\makecell{Sanitization Method}} & \textbf{\makecell{Mauve}} & \textbf{\makecell{Embedding}} & \textbf{\makecell{PII Existence}} & \textbf{\makecell{Lexical Distance}} & \textbf{\makecell{Semantic Distance \\(Ours)}} \\

\midrule
\multirow{10}{*}{\textbf{MedQA}}
& No Sanitization & 0.00$_{\text{(0.00)}}$  & 0.04$_{\text{(0.00)}}$  & 0.00$_{\text{(0.00)}}$  & 0.71$_{\text{(0.00)}}$  & 0.15$_{\text{(0.01)}}$  \\

\cmidrule{2-7}
& Sanitize \& Paraphrase & 0.93$_{\text{(0.01)}}$  & 0.35$_{\text{(0.01)}}$  & 0.80$_{\text{(0.00)}}$  & 0.83$_{\text{(0.00)}}$  & 0.33$_{\text{(0.02)}}$  \\
& Azure AI PII tool & 0.51$_{\text{(0.00)}}$  & 0.18$_{\text{(0.00)}}$  & 0.99$_{\text{(0.00)}}$  & 0.73$_{\text{(0.00)}}$  & 0.26$_{\text{(0.01)}}$  \\
& Span Sanitization & 0.79$_{\text{(0.02)}}$  & 0.38$_{\text{(0.01)}}$  & 0.61$_{\text{(0.01)}}$  & 0.81$_{\text{(0.00)}}$  & 0.55$_{\text{(0.00)}}$  \\
& Iterative Anonymization & 0.74$_{\text{(0.04)}}$  & 0.41$_{\text{(0.00)}}$  & 0.76$_{\text{(0.01)}}$  & 0.81$_{\text{(0.01)}}$  & 0.53$_{\text{(0.01)}}$  \\
& Data Synthesis / $\varepsilon = \infty$ & 0.01$_{\text{(0.01)}}$  & 0.31$_{\text{(0.01)}}$  & 0.15$_{\text{(0.01)}}$  & 0.76$_{\text{(0.01)}}$  & 0.52$_{\text{(0.02)}}$  \\
& DP with $\varepsilon = 1024$ & 0.17$_{\text{(0.02)}}$  & 0.55$_{\text{(0.00)}}$  & 0.89$_{\text{(0.01)}}$  & 0.84$_{\text{(0.00)}}$  & 0.90$_{\text{(0.00)}}$  \\
& DP with $\varepsilon = 64$ & 0.25$_{\text{(0.04)}}$  & 0.56$_{\text{(0.00)}}$  & 0.91$_{\text{(0.00)}}$  & 0.85$_{\text{(0.00)}}$  & 0.91$_{\text{(0.00)}}$  \\
& DP with $\varepsilon = 3$ & 0.38$_{\text{(0.02)}}$  & 0.57$_{\text{(0.00)}}$  & 0.91$_{\text{(0.01)}}$  & 0.85$_{\text{(0.00)}}$  & 0.92$_{\text{(0.00)}}$  \\

\midrule
\multirow{10}{*}{\textbf{WildChat}}
& No Sanitization & 0.00$_{\text{(0.00)}}$  & 0.00$_{\text{(0.00)}}$  & 0.00$_{\text{(0.00)}}$  & 0.54$_{\text{(0.01)}}$  & 0.15$_{\text{(0.00)}}$  \\

\cmidrule{2-7}

& Sanitize \& Paraphrase & 0.80$_{\text{(0.01)}}$  & 0.39$_{\text{(0.00)}}$  & 0.44$_{\text{(0.00)}}$  & 0.74$_{\text{(0.00)}}$  & 0.38$_{\text{(0.01)}}$  \\
& Azure AI PII tool & 0.26$_{\text{(0.00)}}$  & 0.30$_{\text{(0.01)}}$  & 0.70$_{\text{(0.01)}}$  & 0.58$_{\text{(0.00)}}$  & 0.26$_{\text{(0.01)}}$  \\
& Span Sanitization & 0.72$_{\text{(0.00)}}$  & 0.30$_{\text{(0.01)}}$  & 0.10$_{\text{(0.01)}}$  & 0.64$_{\text{(0.00)}}$  & 0.29$_{\text{(0.01)}}$  \\
& Iterative Anonymization & 0.81$_{\text{(0.00)}}$  & 0.44$_{\text{(0.01)}}$  & 0.33$_{\text{(0.02)}}$  & 0.73$_{\text{(0.01)}}$  & 0.46$_{\text{(0.01)}}$  \\
& Data Synthesis / $\varepsilon = \infty$ & 0.95$_{\text{(0.01)}}$  & 0.61$_{\text{(0.01)}}$  & 0.51$_{\text{(0.02)}}$  & 0.88$_{\text{(0.00)}}$  & 0.83$_{\text{(0.00)}}$  \\
& DP with $\varepsilon = 1024$ & 0.87$_{\text{(0.02)}}$  & 0.63$_{\text{(0.00)}}$  & 0.50$_{\text{(0.04)}}$  & 0.89$_{\text{(0.00)}}$  & 0.88$_{\text{(0.01)}}$  \\
& DP with $\varepsilon = 64$ & 0.89$_{\text{(0.01)}}$  & 0.64$_{\text{(0.00)}}$  & 0.51$_{\text{(0.01)}}$  & 0.89$_{\text{(0.00)}}$  & 0.88$_{\text{(0.00)}}$  \\
& DP with $\varepsilon = 3$ & 0.88$_{\text{(0.02)}}$  & 0.65$_{\text{(0.00)}}$  & 0.62$_{\text{(0.00)}}$  & 0.89$_{\text{(0.00)}}$  & 0.88$_{\text{(0.00)}}$  \\

\bottomrule
\end{tabular}
}
\end{table*}

%% file: tables/category_table.tex
\begin{table*}[htbp]
\centering
\caption{Sensitive information categories for classifying privacy leakage types in the dataset.}
\vspace{1ex}
\label{tab:sensitive-category}\adjustbox{max width=0.95\textwidth}{%

\begin{tabular}{p{4cm}p{12cm}}
    \toprule
    \textbf{Category} & \textbf{Description} \\
    \midrule
    Age & Any mention of a person's age, e.g., ``23-year-old'' \\
    Gender & References to gender identity, e.g., ``woman,'' ``non-binary person'' \\
    Sexual\_Orientation & Mentions of sexual orientation, e.g., ``gay couple'' \\
    Race\_Nationality & References to race, ethnicity, or nationality \\
    Spouse & Mentions of a person's wife, husband, or spouse \\
    Partner & References to a person's girlfriend, boyfriend, or partner \\
    Relationship\_Status & Mentions of marital status, being in a romantic relationship, or being single \\
    Family & References to family members or family structures \\
    Health (Only used in WildChat) & Includes a wide range of health-related information, from specific diseases or conditions to medications, medical tests, or treatments \\
    Mental\_Health (Only used in WildChat) & Includes a broad range of emotional states and mental health conditions, from feelings of sadness or anxiety to specific diagnoses\\
    Location & Captures specific geographical details about where a person lives or is located. Includes precise locations such as addresses, cities, countries, or distinctive landmarks \\
    Appearance & Physical descriptions of individuals, e.g., ``He is 6'2'' \\
    Pet & Information about a person's pets or animals \\
    Occupation & References to a person's job or profession \\
    Education & Information about a person's educational background or current studies \\
    Finance & Any details about financial situations or status, not necessarily exact amounts \\
    \midrule
    \textbf{MedQA Specific} \\
    \midrule
    Chief\_Concern & The primary reason for a medical visit or the main health issue \\
    History\_of\_Present\_Illness & Detailed account of the development of the current health problem \\
    Past\_Medical\_History & Previous illnesses, surgeries, or significant health events \\
    Medications & Current or past medications, including dosages and frequencies \\
    Allergies\_Reactions & Any known allergies or adverse reactions to medications or substances \\
    Social\_History & Information about lifestyle, habits, occupation, and living situation that may impact health \\
    Family\_History & Health information about immediate family members \\
    Review\_of\_Systems & Systematic review of body systems for additional symptoms \\
    Physical\_Exam & Findings from a physical examination \\
    Diagnostic\_Results & Results from laboratory tests (blood, urine, etc.), radiologic studies (X-rays, CT scans, MRIs, etc.), and other diagnostic procedures (e.g., EKG interpretations) \\
    \bottomrule
    \hline
\end{tabular}}
\end{table*}

%% file: tables/examples_table.tex
\begin{table*}[t]
\centering
\tiny
\caption{Comparison of original and re-identified records from the MedQA dataset sanitized with the data synthesis sanitization method, along with corresponding matching claims. We demonstrate the attributes extracted through our inference method. 
}
\vspace{1ex}
\begin{tabular}{p{0.2\textwidth}p{0.2\textwidth}p{0.2\textwidth}p{0.2\textwidth}}
\toprule
\textbf{Original Record} & \textbf{Our Method Match} & \textbf{Claims Used for Matching} & \textbf{Privacy Leaks Detected by Semantic Similarity}  \\
\midrule
A 23-year-old woman is brought to the emergency department ... She says that she feels "empty inside" and has been hearing voices telling her that she is worthless. ... She does not drink alcohol or use illicit drugs. ... On mental status examination, her speech is slow and monotonous; she abruptly stops talking in the middle of sentences and does not finish them. She occasionally directs her attention to the ceiling as if she were listening to someone. & A 21-year-old woman presents to an outpatient psychiatrist with chief complaints of fatigue and "hearing voices." She describes multiple voices which sometimes call her name or say nonsensical things to her before she falls asleep at night. ... The patient has no significant past medical or psychiatric history. She does not smoke or drink alcohol. ... & She abruptly stops talking in the middle of sentences.\newline
She does not finish her sentences.\newline
She occasionally directs her attention to the ceiling as if she were listening to someone. & 1. Young adult (early 20s)\newline
2. Presence of auditory hallucinations\newline
3. No substance use history\newline
4. Potential psychotic disorder \\
\midrule
A 34-year-old woman, gravida 1, para 0, at 16 weeks' gestation comes to the physician for a routine prenatal visit. ... Serum studies show:\newline
Alpha-fetoprotein decreased\newline
Unconjugated estriol decreased\newline
Human chorionic gonadotropin increased\newline
Inhibin A increased & A 26-year-old primigravid woman comes to the physician ... for her first prenatal visit. ... Maternal serum studies show low $\alpha$-fetoprotein and free estriol concentrations, and increased inhibin A and $\beta$-human chorionic gonadotropin concentrations. & Serum human chorionic gonadotropin levels are increased.\newline
Serum inhibin A levels are increased.\newline
The patient wants a definitive diagnosis as quickly as possible. & 1. Pregnant woman\newline
2. First pregnancy\newline
3. Abnormal serum markers\newline
4. Potential fetal abnormality  \\
\midrule
A 58-year-old chronic smoker known to have chronic bronchitis for the last 20 years presents to his physician ... Right heart catheterization is performed, which indicates a pulmonary artery pressure of 30 mm Hg and a pulmonary capillary wedge pressure of 13 mm Hg. There is a significant drop in pulmonary artery pressure after the administration of inhaled nitric oxide.& A 51-year-old man comes to the physician because of progressively worsening dyspnea on exertion and fatigue for the past 2 months. ... Coarse crackles are heard at the lung bases bilaterally. ... An x-ray of the chest shows globular enlargement of the cardiac shadow with prominent hila and bilateral fluffy infiltrates. ... & Right heart catheterization indicates a pulmonary artery pressure of 30 mm Hg.\newline
Right heart catheterization indicates a pulmonary capillary wedge pressure of 13 mm Hg.\newline
There is a significant drop in pulmonary artery pressure after the administration of inhaled nitric oxide. & 1. Middle-aged man\newline
2. Progressive breathing difficulty\newline
3. Indication of lung disease\newline
4. Potential heart involvement \\
\midrule
A 56-year-old woman comes to the emergency department because of worsening pain and swelling in her right knee for 3 days. She underwent a total knee arthroplasty of her right knee joint 5 months ago. ... Analysis of the synovial fluid shows: ... WBC count 78,000/mm3\newline
Segmented neutrophils 94\%\newline
Lymphocytes 6\%\newline
Synovial fluid is sent for culture and antibiotic sensitivity. & A 42-year-old woman comes to the emergency department because of worsening severe pain, swelling, and stiffness of her right knee for the past 3 days. ... Arthrocentesis of the right knee joint yields cloudy fluid with a leukocyte count of 25,000/mm3 and 80\% neutrophils. ... & Analysis of the synovial fluid shows lymphocytes 6\%.\newline
Synovial fluid is sent for culture.\newline
Synovial fluid is sent for antibiotic sensitivity. & 1. Middle-aged woman\newline
2. Right knee problem\newline
3. Joint inflammation\newline
4. Potential infection \\
\bottomrule
\end{tabular}
\label{tab:examples}
\end{table*}

%% file: arxiv-final.bbl
\begin{thebibliography}{59}
\providecommand{\natexlab}[1]{#1}
\providecommand{\url}[1]{#1}
\csname url@samestyle\endcsname
\providecommand{\newblock}{\relax}
\providecommand{\bibinfo}[2]{#2}
\providecommand{\BIBentrySTDinterwordspacing}{\spaceskip=0pt\relax}
\providecommand{\BIBentryALTinterwordstretchfactor}{4}
\providecommand{\BIBentryALTinterwordspacing}{\spaceskip=\fontdimen2\font plus
\BIBentryALTinterwordstretchfactor\fontdimen3\font minus \fontdimen4\font\relax}
\providecommand{\BIBforeignlanguage}[2]{{%
\expandafter\ifx\csname l@#1\endcsname\relax
\typeout{** WARNING: IEEEtranN.bst: No hyphenation pattern has been}%
\typeout{** loaded for the language `#1'. Using the pattern for}%
\typeout{** the default language instead.}%
\else
\language=\csname l@#1\endcsname
\fi
#2}}
\providecommand{\BIBdecl}{\relax}
\BIBdecl

\bibitem[{Federal Data Strategy}(2020)]{FDS20}
\BIBentryALTinterwordspacing
{Federal Data Strategy}, ``Federal data strategy,'' 2020, accessed 2024-09-01. [Online]. Available: \url{https://strategy.data.gov/}
\BIBentrySTDinterwordspacing

\bibitem[McMahan et~al.(2017)McMahan, Moore, Ramage, Hampson, and Ag{\"u}era~y Arcas]{mcmahan2017communication}
\BIBentryALTinterwordspacing
B.~McMahan, E.~Moore, D.~Ramage, S.~Hampson, and B.~Ag{\"u}era~y Arcas, ``Communication-efficient learning of deep networks from decentralized data,'' in \emph{International Conference on Artificial Intelligence and Statistics (AISTATS)}, 2017, pp. 1273--1282. [Online]. Available: \url{http://proceedings.mlr.press/v54/mcmahan17a.html}
\BIBentrySTDinterwordspacing

\bibitem[Narayanan and Shmatikov(2006)]{narayanan2006break}
A.~Narayanan and V.~Shmatikov, ``How to break anonymity of the netflix prize dataset,'' \emph{arXiv preprint cs/0610105}, 2006.

\bibitem[Abowd et~al.(2023)Abowd, Adams, Ashmead, Darais, Dey, Garfinkel, Goldschlag, Kifer, Leclerc, Lew, et~al.]{abowd20232010}
J.~M. Abowd, T.~Adams, R.~Ashmead, D.~Darais, S.~Dey, S.~L. Garfinkel, N.~Goldschlag, D.~Kifer, P.~Leclerc, E.~Lew \emph{et~al.}, ``The 2010 census confidentiality protections failed, here's how and why,'' National Bureau of Economic Research, Tech. Rep., 2023.

\bibitem[Yan et~al.(2024)Yan, Li, Xu, Dong, Zhang, Ren, and Cheng]{yan2024protecting}
B.~Yan, K.~Li, M.~Xu, Y.~Dong, Y.~Zhang, Z.~Ren, and X.~Cheng, ``On protecting the data privacy of large language models (llms): A survey,'' \emph{arXiv preprint arXiv:2403.05156}, 2024.

\bibitem[Ganta et~al.(2008)Ganta, Kasiviswanathan, and Smith]{ganta2008composition}
S.~R. Ganta, S.~P. Kasiviswanathan, and A.~Smith, ``Composition attacks and auxiliary information in data privacy,'' in \emph{Proceedings of the 14th ACM SIGKDD international conference on Knowledge discovery and data mining}, 2008, pp. 265--273.

\bibitem[Venkatesan et~al.(2008)Venkatesan, Gupta, Roy, and Mohania]{venkatesan2008efficient}
T.~Venkatesan, H.~Gupta, P.~Roy, and M.~K. Mohania, ``Efficient techniques for document sanitization,'' in \emph{ACM Conference on Information and Knowledge Management}, 2008.

\bibitem[Lison et~al.(2021)Lison, Pil{\'a}n, Sanchez, Batet, and {\O}vrelid]{lison-etal-2021-anonymisation}
\BIBentryALTinterwordspacing
P.~Lison, I.~Pil{\'a}n, D.~Sanchez, M.~Batet, and L.~{\O}vrelid, ``Anonymisation models for text data: State of the art, challenges and future directions,'' in \emph{Proceedings of the 59th Annual Meeting of the Association for Computational Linguistics and the 11th International Joint Conference on Natural Language Processing (Volume 1: Long Papers)}, C.~Zong, F.~Xia, W.~Li, and R.~Navigli, Eds.\hskip 1em plus 0.5em minus 0.4em\relax Online: Association for Computational Linguistics, Aug. 2021, pp. 4188--4203. [Online]. Available: \url{https://aclanthology.org/2021.acl-long.323/}
\BIBentrySTDinterwordspacing

\bibitem[Anandan and Clifton(2011)]{6040853}
B.~Anandan and C.~Clifton, ``Significance of term relationships on anonymization,'' in \emph{2011 IEEE/WIC/ACM International Conferences on Web Intelligence and Intelligent Agent Technology}, vol.~3, 2011, pp. 253--256.

\bibitem[Pil{\'a}n et~al.(2022)Pil{\'a}n, Lison, {\O}vrelid, Papadopoulou, S{\'a}nchez, and Batet]{pilan2022text}
I.~Pil{\'a}n, P.~Lison, L.~{\O}vrelid, A.~Papadopoulou, D.~S{\'a}nchez, and M.~Batet, ``The text anonymization benchmark (tab): A dedicated corpus and evaluation framework for text anonymization,'' \emph{Computational Linguistics}, vol.~48, no.~4, pp. 1053--1101, 2022.

\bibitem[Boutet et~al.(2025)Boutet, Kazdam, Magnana, and Zimmermann]{boutet2025anonymization}
A.~Boutet, Z.~E. Kazdam, L.~Magnana, and H.~Zimmermann, ``Anonymization by design of language modeling,'' \emph{arXiv preprint arXiv:2501.02407}, 2025.

\bibitem[Carlini et~al.(2019)Carlini, Liu, Erlingsson, Kos, and Song]{carlini2019secret}
N.~Carlini, C.~Liu, {\'U}.~Erlingsson, J.~Kos, and D.~Song, ``The secret sharer: Evaluating and testing unintended memorization in neural networks,'' in \emph{28th USENIX Security Symposium (USENIX Security 19)}, 2019, pp. 267--284.

\bibitem[Carlini et~al.(2022)Carlini, Ippolito, Jagielski, Lee, Tramer, and Zhang]{carlini2022quantifying}
N.~Carlini, D.~Ippolito, M.~Jagielski, K.~Lee, F.~Tramer, and C.~Zhang, ``Quantifying memorization across neural language models,'' in \emph{The Eleventh International Conference on Learning Representations}, 2022.

\bibitem[Huang et~al.(2023)Huang, Gupta, Zhong, Li, and Chen]{huang2023privacy}
Y.~Huang, S.~Gupta, Z.~Zhong, K.~Li, and D.~Chen, ``Privacy implications of retrieval-based language models,'' \emph{arXiv preprint arXiv:2305.14888}, 2023.

\bibitem[Jin et~al.(2021)Jin, Pan, Oufattole, Weng, Fang, and Szolovits]{jin2021disease}
D.~Jin, E.~Pan, N.~Oufattole, W.-H. Weng, H.~Fang, and P.~Szolovits, ``What disease does this patient have? a large-scale open domain question answering dataset from medical exams,'' \emph{Applied Sciences}, vol.~11, no.~14, p. 6421, 2021.

\bibitem[Zhao et~al.(2024)Zhao, Ren, Hessel, Cardie, Choi, and Deng]{zhao2024wildchat}
\BIBentryALTinterwordspacing
W.~Zhao, X.~Ren, J.~Hessel, C.~Cardie, Y.~Choi, and Y.~Deng, ``Wildchat: 1m chat{GPT} interaction logs in the wild,'' in \emph{The Twelfth International Conference on Learning Representations}, 2024. [Online]. Available: \url{https://openreview.net/forum?id=Bl8u7ZRlbM}
\BIBentrySTDinterwordspacing

\bibitem[Mireshghallah et~al.(2024)Mireshghallah, Antoniak, More, Choi, and Farnadi]{mireshghallah2024trust}
N.~Mireshghallah, M.~Antoniak, Y.~More, Y.~Choi, and G.~Farnadi, ``{Trust} {No} {Bot}: {Discovering} {Personal} {Disclosures} in {Human}-{LLM} {Conversations} in the {Wild},'' in \emph{The First Conference on Language Modeling ({COLM})}, October 2024.

\bibitem[Staab et~al.(2024)Staab, Vero, Balunovi{\'c}, and Vechev]{staab2024large}
R.~Staab, M.~Vero, M.~Balunovi{\'c}, and M.~Vechev, ``Large language models are advanced anonymizers,'' \emph{arXiv preprint arXiv:2402.13846}, 2024.

\bibitem[Dou et~al.(2024)Dou, Krsek, Naous, Kabra, Das, Ritter, and Xu]{dou2023reducing}
\BIBentryALTinterwordspacing
Y.~Dou, I.~Krsek, T.~Naous, A.~Kabra, S.~Das, A.~Ritter, and W.~Xu, ``Reducing privacy risks in online self-disclosures with language models,'' in \emph{Proceedings of the 62nd Annual Meeting of the Association for Computational Linguistics (Volume 1: Long Papers)}, L.-W. Ku, A.~Martins, and V.~Srikumar, Eds.\hskip 1em plus 0.5em minus 0.4em\relax Bangkok, Thailand: Association for Computational Linguistics, Aug. 2024, pp. 13\,732--13\,754. [Online]. Available: \url{https://aclanthology.org/2024.acl-long.741}
\BIBentrySTDinterwordspacing

\bibitem[Yue et~al.(2023)Yue, Inan, Li, Kumar, McAnallen, Shajari, Sun, Levitan, and Sim]{yue2023synthetic}
X.~Yue, H.~Inan, X.~Li, G.~Kumar, J.~McAnallen, H.~Shajari, H.~Sun, D.~Levitan, and R.~Sim, ``Synthetic text generation with differential privacy: A simple and practical recipe,'' in \emph{Proceedings of the 61st Annual Meeting of the Association for Computational Linguistics (Volume 1: Long Papers)}, 2023, pp. 1321--1342.

\bibitem[Hundepool et~al.(2012)Hundepool, Domingo-Ferrer, Franconi, Giessing, Nordholt, Spicer, and De~Wolf]{hundepool2012statistical}
A.~Hundepool, J.~Domingo-Ferrer, L.~Franconi, S.~Giessing, E.~S. Nordholt, K.~Spicer, and P.-P. De~Wolf, \emph{Statistical disclosure control}.\hskip 1em plus 0.5em minus 0.4em\relax John Wiley \& Sons, 2012.

\bibitem[Bellovin et~al.(2019)Bellovin, Dutta, and Reitinger]{bellovin2019privacy}
S.~M. Bellovin, P.~K. Dutta, and N.~Reitinger, ``Privacy and synthetic datasets,'' \emph{Stan. Tech. L. Rev.}, vol.~22, p.~1, 2019.

\bibitem[Garfinkel(2015)]{garfinkel2015deidentification}
\BIBentryALTinterwordspacing
S.~L. Garfinkel, ``De-identification of personal information,'' National Institute of Standards and Technology, NISTIR 8053, 2015, this publication is available free of charge. [Online]. Available: \url{http://dx.doi.org/10.6028/NIST.IR.8053}
\BIBentrySTDinterwordspacing

\bibitem[Giuffr{\`e} and Shung(2023)]{giuffre2023harnessing}
\BIBentryALTinterwordspacing
M.~Giuffr{\`e} and D.~L. Shung, ``Harnessing the power of synthetic data in healthcare: innovation, application, and privacy,'' \emph{npj Digital Medicine}, vol.~6, p. 186, 2023, received: 15 April 2023, Accepted: 14 September 2023, Published: 09 October 2023. [Online]. Available: \url{https://doi.org/10.1038/s41746-023-00927-3}
\BIBentrySTDinterwordspacing

\bibitem[El~Emam et~al.(2011)El~Emam, Jonker, Arbuckle, and Malin]{el2011systematic}
K.~El~Emam, E.~Jonker, L.~Arbuckle, and B.~Malin, ``A systematic review of re-identification attacks on health data,'' \emph{PloS one}, vol.~6, no.~12, p. e28071, 2011.

\bibitem[Carlini et~al.(2021)Carlini, Tramer, Wallace, Jagielski, Herbert-Voss, Lee, Roberts, Brown, Song, Erlingsson, et~al.]{carlini2021extracting}
N.~Carlini, F.~Tramer, E.~Wallace, M.~Jagielski, A.~Herbert-Voss, K.~Lee, A.~Roberts, T.~Brown, D.~Song, U.~Erlingsson \emph{et~al.}, ``Extracting training data from large language models,'' in \emph{30th USENIX Security Symposium (USENIX Security 21)}, 2021, pp. 2633--2650.

\bibitem[Stadler et~al.(2022)Stadler, Oprisanu, and Troncoso]{stadler2022synthetic}
T.~Stadler, B.~Oprisanu, and C.~Troncoso, ``Synthetic data--anonymisation groundhog day,'' in \emph{31st USENIX Security Symposium (USENIX Security 22)}, 2022, pp. 1451--1468.

\bibitem[Yale et~al.(2019)Yale, Dash, Dutta, Guyon, Pavao, and Bennett]{yale2019assessing}
A.~Yale, S.~Dash, R.~Dutta, I.~Guyon, A.~Pavao, and K.~P. Bennett, ``Assessing privacy and quality of synthetic health data,'' in \emph{Proceedings of the Conference on Artificial Intelligence for Data Discovery and Reuse}, 2019, pp. 1--4.

\bibitem[Annamalai et~al.(2024)Annamalai, Gadotti, and Rocher]{annamalai2024linear}
M.~S. M.~S. Annamalai, A.~Gadotti, and L.~Rocher, ``A linear reconstruction approach for attribute inference attacks against synthetic data,'' in \emph{USENIX Association}, 2024.

\bibitem[Mendels et~al.(2018)Mendels, Peled, Vaisman~Levy, Hart, Rosenthal, Lahiani, et~al.]{MsPresidio}
\BIBentryALTinterwordspacing
O.~Mendels, C.~Peled, N.~Vaisman~Levy, S.~Hart, T.~Rosenthal, L.~Lahiani \emph{et~al.}, ``{Microsoft Presidio}: Context aware, pluggable and customizable pii anonymization service for text and images,'' Microsoft, 2018. [Online]. Available: \url{https://microsoft.github.io/presidio}
\BIBentrySTDinterwordspacing

\bibitem[Montani et~al.(2022)Montani, Honnibal, Honnibal, Landeghem, Boyd, Peters, McCann, Samsonov, Geovedi, O'Regan, Altinok, Orosz, Kristiansen, Roman, Bot, Miranda, Fiedler, de~Kok, Howard, Edward, Phatthiyaphaibun, Tamura, Bozek, murat, Amery, Daniels, Böing, Tippa, and Baumgartner]{ines_montani_2022_6397450}
\BIBentryALTinterwordspacing
I.~Montani, M.~Honnibal, M.~Honnibal, S.~V. Landeghem, A.~Boyd, H.~Peters, P.~O. McCann, M.~Samsonov, J.~Geovedi, J.~O'Regan, D.~Altinok, G.~Orosz, S.~L. Kristiansen, Roman, E.~Bot, L.~Miranda, L.~Fiedler, D.~de~Kok, G.~Howard, Edward, W.~Phatthiyaphaibun, Y.~Tamura, S.~Bozek, murat, M.~Amery, R.~Daniels, B.~Böing, P.~K. Tippa, and P.~Baumgartner, ``{explosion/spaCy: v3.1.6: Workaround for Click/Typer issues},'' Mar. 2022. [Online]. Available: \url{https://doi.org/10.5281/zenodo.6397450}
\BIBentrySTDinterwordspacing

\bibitem[Zhou et~al.(2024)Zhou, Xu, Wu, and Li]{zhou2024rescribersmallerllmpowereduserleddata}
\BIBentryALTinterwordspacing
J.~Zhou, E.~Xu, Y.~Wu, and T.~Li, ``Rescriber: Smaller-llm-powered user-led data minimization for navigating privacy trade-offs in llm-based conversational agent,'' 2024. [Online]. Available: \url{https://arxiv.org/abs/2410.11876}
\BIBentrySTDinterwordspacing

\bibitem[Morris et~al.(2022)Morris, Chiu, Zabih, and Rush]{morrisUnsupervisedTextDeidentification2022}
\BIBentryALTinterwordspacing
J.~Morris, J.~Chiu, R.~Zabih, and A.~Rush, ``Unsupervised text deidentification,'' in \emph{Findings of the Association for Computational Linguistics: EMNLP 2022}, Y.~Goldberg, Z.~Kozareva, and Y.~Zhang, Eds.\hskip 1em plus 0.5em minus 0.4em\relax Abu Dhabi, United Arab Emirates: Association for Computational Linguistics, Dec. 2022, pp. 4777--4788. [Online]. Available: \url{https://aclanthology.org/2022.findings-emnlp.352}
\BIBentrySTDinterwordspacing

\bibitem[Liu et~al.(2024)Liu, Wei, Liu, Si, Zhang, Rao, Zheng, Peng, Yang, Zhou, et~al.]{liu2024best}
R.~Liu, J.~Wei, F.~Liu, C.~Si, Y.~Zhang, J.~Rao, S.~Zheng, D.~Peng, D.~Yang, D.~Zhou \emph{et~al.}, ``Best practices and lessons learned on synthetic data,'' \emph{arXiv preprint arXiv:2404.07503}, 2024.

\bibitem[Xie et~al.(2018)Xie, Lin, Wang, Wang, and Zhou]{xie2018differentially}
L.~Xie, K.~Lin, S.~Wang, F.~Wang, and J.~Zhou, ``Differentially private generative adversarial network,'' \emph{arXiv preprint arXiv:1802.06739}, 2018.

\bibitem[Torkzadehmahani et~al.(2019)Torkzadehmahani, Kairouz, and Paten]{Torkzadehmahani2019DPCGANDP}
\BIBentryALTinterwordspacing
R.~Torkzadehmahani, P.~Kairouz, and B.~Paten, ``Dp-cgan: Differentially private synthetic data and label generation,'' \emph{2019 IEEE/CVF Conference on Computer Vision and Pattern Recognition Workshops (CVPRW)}, pp. 98--104, 2019. [Online]. Available: \url{https://api.semanticscholar.org/CorpusID:198181039}
\BIBentrySTDinterwordspacing

\bibitem[Weggenmann et~al.(2022)Weggenmann, Rublack, Andrejczuk, Mattern, and Kerschbaum]{weggenmann_dp-vae_2022}
\BIBentryALTinterwordspacing
B.~Weggenmann, V.~Rublack, M.~Andrejczuk, J.~Mattern, and F.~Kerschbaum, ``{DP}-{VAE}: Human-readable text anonymization for online reviews with differentially private variational autoencoders,'' in \emph{Proceedings of the {ACM} Web Conference 2022}.\hskip 1em plus 0.5em minus 0.4em\relax {ACM}, 2022, pp. 721--731. [Online]. Available: \url{https://dl.acm.org/doi/10.1145/3485447.3512232}
\BIBentrySTDinterwordspacing

\bibitem[Igamberdiev and Habernal(2023)]{igamberdiev_dp-bart_2023}
\BIBentryALTinterwordspacing
T.~Igamberdiev and I.~Habernal, ``{DP}-{BART} for privatized text rewriting under local differential privacy,'' 2023. [Online]. Available: \url{http://arxiv.org/abs/2302.07636}
\BIBentrySTDinterwordspacing

\bibitem[Bo et~al.(2021)Bo, Ding, Fung, and Iqbal]{bo_er-ae_2021}
\BIBentryALTinterwordspacing
H.~Bo, S.~H.~H. Ding, B.~C.~M. Fung, and F.~Iqbal, ``{ER}-{AE}: Differentially private text generation for authorship anonymization,'' in \emph{Proceedings of the 2021 Conference of the North American Chapter of the Association for Computational Linguistics: Human Language Technologies}.\hskip 1em plus 0.5em minus 0.4em\relax Association for Computational Linguistics, 2021, pp. 3997--4007. [Online]. Available: \url{https://aclanthology.org/2021.naacl-main.314}
\BIBentrySTDinterwordspacing

\bibitem[Igamberdiev et~al.(2022)Igamberdiev, Arnold, and Habernal]{Igamberdiev.2022.COLING}
T.~Igamberdiev, T.~Arnold, and I.~Habernal, ``Dp-rewrite: Towards reproducibility and transparency in differentially private text rewriting,'' in \emph{Proceedings of the 29th International Conference on Computational Linguistics}.\hskip 1em plus 0.5em minus 0.4em\relax Gyeongju, Republic of Korea: International Committee on Computational Linguistics, 2022, p. (to appear).

\bibitem[Mattern et~al.(2022)Mattern, Jin, Weggenmann, Schoelkopf, and Sachan]{mattern_differentially_2022}
\BIBentryALTinterwordspacing
J.~Mattern, Z.~Jin, B.~Weggenmann, B.~Schoelkopf, and M.~Sachan, ``Differentially private language models for secure data sharing,'' 2022. [Online]. Available: \url{http://arxiv.org/abs/2210.13918}
\BIBentrySTDinterwordspacing

\bibitem[Mireshghallah et~al.(2022)Mireshghallah, Su, Hashimoto, Eisner, and Shin]{mireshghallah2022privacy}
F.~Mireshghallah, Y.~Su, T.~Hashimoto, J.~Eisner, and R.~Shin, ``Privacy-preserving domain adaptation of semantic parsers,'' \emph{arXiv preprint arXiv:2212.10520}, 2022.

\bibitem[Kurakin et~al.(2023)Kurakin, Ponomareva, Syed, MacDermed, and Terzis]{kurakin2023harnessing}
A.~Kurakin, N.~Ponomareva, U.~Syed, L.~MacDermed, and A.~Terzis, ``Harnessing large-language models to generate private synthetic text,'' \emph{arXiv preprint arXiv:2306.01684}, 2023.

\bibitem[Pang et~al.(2024)Pang, Lu, Wang, Fu, Zhou, Xue, and Li]{pang2024reconstructiondifferentiallyprivatetext}
\BIBentryALTinterwordspacing
S.~Pang, Z.~Lu, H.~Wang, P.~Fu, Y.~Zhou, M.~Xue, and B.~Li, ``Reconstruction of differentially private text sanitization via large language models,'' 2024. [Online]. Available: \url{https://arxiv.org/abs/2410.12443}
\BIBentrySTDinterwordspacing

\bibitem[Morris et~al.(2024)Morris, Campion, Nutheti, Peng, Raj, Zabih, and Cole]{morris2024diriadversarialpatientreidentification}
\BIBentryALTinterwordspacing
J.~X. Morris, T.~R. Campion, S.~L. Nutheti, Y.~Peng, A.~Raj, R.~Zabih, and C.~L. Cole, ``Diri: Adversarial patient reidentification with large language models for evaluating clinical text anonymization,'' 2024. [Online]. Available: \url{https://arxiv.org/abs/2410.17035}
\BIBentrySTDinterwordspacing

\bibitem[Ramesh et~al.(2024)Ramesh, Gandhi, Madaan, Bauer, Peris, and Field]{ramesh2024evaluatingdifferentiallyprivatesynthetic}
\BIBentryALTinterwordspacing
K.~Ramesh, N.~Gandhi, P.~Madaan, L.~Bauer, C.~Peris, and A.~Field, ``Evaluating differentially private synthetic data generation in high-stakes domains,'' 2024. [Online]. Available: \url{https://arxiv.org/abs/2410.08327}
\BIBentrySTDinterwordspacing

\bibitem[Min et~al.(2023)Min, Krishna, Lyu, Lewis, Yih, Koh, Iyyer, Zettlemoyer, and Hajishirzi]{minFActScoreFinegrainedAtomic2023b}
\BIBentryALTinterwordspacing
S.~Min, K.~Krishna, X.~Lyu, M.~Lewis, W.-t. Yih, P.~Koh, M.~Iyyer, L.~Zettlemoyer, and H.~Hajishirzi, ``{FA}ct{S}core: Fine-grained atomic evaluation of factual precision in long form text generation,'' in \emph{Proceedings of the 2023 Conference on Empirical Methods in Natural Language Processing}, H.~Bouamor, J.~Pino, and K.~Bali, Eds.\hskip 1em plus 0.5em minus 0.4em\relax Singapore: Association for Computational Linguistics, Dec. 2023, pp. 12\,076--12\,100. [Online]. Available: \url{https://aclanthology.org/2023.emnlp-main.741}
\BIBentrySTDinterwordspacing

\bibitem[Dubey et~al.(2024)Dubey, Jauhri, et~al.]{dubey_llama_2024}
\BIBentryALTinterwordspacing
A.~Dubey, A.~Jauhri \emph{et~al.}, ``The llama 3 herd of models,'' 2024. [Online]. Available: \url{http://arxiv.org/abs/2407.21783}
\BIBentrySTDinterwordspacing

\bibitem[Muennighoff et~al.(2024)Muennighoff, Su, Wang, Yang, Wei, Yu, Singh, and Kiela]{muennighoff2024generative}
N.~Muennighoff, H.~Su, L.~Wang, N.~Yang, F.~Wei, T.~Yu, A.~Singh, and D.~Kiela, ``Generative representational instruction tuning,'' \emph{arXiv preprint arXiv:2402.09906}, 2024.

\bibitem[Lin(2004)]{lin2004rouge}
C.-Y. Lin, ``Rouge: A package for automatic evaluation of summaries,'' in \emph{Text summarization branches out}, 2004, pp. 74--81.

\bibitem[Xiao et~al.(2024)Xiao, Jin, Bai, Wu, Yang, Luo, Yu, Zhao, Liu, Gu, et~al.]{xiao2024large}
Y.~Xiao, Y.~Jin, Y.~Bai, Y.~Wu, X.~Yang, X.~Luo, W.~Yu, X.~Zhao, Y.~Liu, Q.~Gu \emph{et~al.}, ``Large language models can be contextual privacy protection learners,'' in \emph{Proceedings of the 2024 Conference on Empirical Methods in Natural Language Processing}, 2024, pp. 14\,179--14\,201.

\bibitem[Frikha et~al.(2024)Frikha, Walha, Nakka, Mendes, Jiang, and Zhou]{frikha2024incognitext}
A.~Frikha, N.~Walha, K.~K. Nakka, R.~Mendes, X.~Jiang, and X.~Zhou, ``Incognitext: Privacy-enhancing conditional text anonymization via llm-based private attribute randomization,'' \emph{arXiv preprint arXiv:2407.02956}, 2024.

\bibitem[Zeng et~al.(2024{\natexlab{a}})Zeng, Zhang, He, Liu, Xing, Xu, Ren, Chang, Wang, Yin, and Tang]{zeng_good_2024}
\BIBentryALTinterwordspacing
S.~Zeng, J.~Zhang, P.~He, Y.~Liu, Y.~Xing, H.~Xu, J.~Ren, Y.~Chang, S.~Wang, D.~Yin, and J.~Tang, ``The good and the bad: Exploring privacy issues in retrieval-augmented generation ({RAG}),'' in \emph{Findings of the Association for Computational Linguistics {ACL} 2024}.\hskip 1em plus 0.5em minus 0.4em\relax Association for Computational Linguistics, 2024, pp. 4505--4524. [Online]. Available: \url{https://aclanthology.org/2024.findings-acl.267}
\BIBentrySTDinterwordspacing

\bibitem[Chiang and Lee(2023)]{chiang_can_2023}
\BIBentryALTinterwordspacing
C.-H. Chiang and H.-y. Lee, ``Can large language models be an alternative to human evaluations?'' in \emph{Proceedings of the 61st Annual Meeting of the Association for Computational Linguistics (Volume 1: Long Papers)}, A.~Rogers, J.~Boyd-Graber, and N.~Okazaki, Eds.\hskip 1em plus 0.5em minus 0.4em\relax Association for Computational Linguistics, 2023, pp. 15\,607--15\,631. [Online]. Available: \url{https://aclanthology.org/2023.acl-long.870}
\BIBentrySTDinterwordspacing

\bibitem[Zeng et~al.(2024{\natexlab{b}})Zeng, Zhang, He, Ren, Zheng, Lu, Xu, Liu, Xing, and Tang]{zeng_mitigating_2024}
S.~Zeng, J.~Zhang, P.~He, J.~Ren, T.~Zheng, H.~Lu, H.~Xu, H.~Liu, Y.~Xing, and J.~Tang, ``Mitigating the privacy issues in retrieval-augmented generation (rag) via pure synthetic data,'' \emph{arXiv preprint arXiv:2406.14773}, 2024.

\bibitem[Lin et~al.(2021)Lin, Ma, Lin, Yang, Pradeep, and Nogueira]{Lin_etal_SIGIR2021_Pyserini}
J.~Lin, X.~Ma, S.-C. Lin, J.-H. Yang, R.~Pradeep, and R.~Nogueira, ``{Pyserini}: A {Python} toolkit for reproducible information retrieval research with sparse and dense representations,'' in \emph{Proceedings of the 44th Annual International ACM SIGIR Conference on Research and Development in Information Retrieval (SIGIR 2021)}, 2021, pp. 2356--2362.

\bibitem[Pillutla et~al.(2021)Pillutla, Swayamdipta, Zellers, Thickstun, Welleck, Choi, and Harchaoui]{pillutla2021mauve}
K.~Pillutla, S.~Swayamdipta, R.~Zellers, J.~Thickstun, S.~Welleck, Y.~Choi, and Z.~Harchaoui, ``Mauve: Measuring the gap between neural text and human text using divergence frontiers,'' \emph{Advances in Neural Information Processing Systems}, vol.~34, pp. 4816--4828, 2021.

\bibitem[Wang et~al.(2020)Wang, Wei, Dong, Bao, Yang, and Zhou]{wang2020minilm}
W.~Wang, F.~Wei, L.~Dong, H.~Bao, N.~Yang, and M.~Zhou, ``Minilm: Deep self-attention distillation for task-agnostic compression of pre-trained transformers,'' \emph{Advances in Neural Information Processing Systems}, vol.~33, pp. 5776--5788, 2020.

\bibitem[Goldberg()]{goldbergpratical}
\BIBentryALTinterwordspacing
C.~Goldberg. {UC} san diego's practical guide to clinical medicine. [Online]. Available: \url{https://meded.ucsd.edu/clinicalmed/write.html}
\BIBentrySTDinterwordspacing

\end{thebibliography}
